\documentclass{article}
\usepackage[utf8]{inputenc}
\usepackage[english]{babel}
\usepackage{url}
\usepackage{hyperref}
\usepackage{amsthm}
\usepackage{amsmath}
\usepackage{amssymb}
\usepackage{dsfont}
\usepackage{tikz}
\usepackage{comment}

\usetikzlibrary{arrows.meta,matrix}

\usepackage{algorithm2e}

\usepackage{xcolor}
\usepackage{authblk}
\usepackage{todonotes}
\usepackage[T1]{fontenc}

\title{On the Parameterized Complexity of Relaxations of Clique}
\author[$*$]{Ambroise Baril}
\author[$\dagger$, $\ddag$]{Antoine Castillon}
\author[$\ddag$]{Nacim Oijid}

\date{} 

\affil[$*$]{Université de Lorraine, CNRS, LORIA, France}
\affil[$\dagger$]{Univ. Lille, CNRS, Centrale Lille, UMR 9189 CRIStAL, F-59000 Lille, France}
\affil[$\ddag$]{Univ. Lyon, Universit\'e Lyon 1, LIRIS UMR CNRS 5205, F-69621, Lyon, France.}

\theoremstyle{definition}
\newtheorem*{pb}{Problem}
\newtheorem{definition}{Definition}

\newtheorem{notation}[definition]{Notation}

\theoremstyle{plain}
\newtheorem{thm}[definition]{Theorem}
\newtheorem{lem}[definition]{Lemma}
\newtheorem{rem}[definition]{Remark}
\newtheorem{claim}[definition]{Claim}

\newtheorem{cor}[definition]{Corollary}

\newcommand{\clique}{{\sc Clique}}
\newcommand{\sclique}{$s$-{\sc Clique}}
\newcommand{\club}{{\sc Club}}
\newcommand{\sclub}{$s$-{\sc Club}}
\newcommand{\gcs}{$\gamma$-{\sc Complete Subgraph}}

\newcommand{\qedclaim}{\hfill $\diamond$ \medskip}
\newenvironment{proofclaim}{\noindent{\em Proof of the claim.}}{\qedclaim}
\newenvironment{proofclaim2}{\noindent{\em Proof of the claim.}}{}

\renewcommand{\deg}{\operatorname{deg}}
\newcommand{\dist}{\operatorname{dist}}
\newcommand{\diam}{\operatorname{diam}}

\begin{document}

\definecolor{vert}{HTML}{008800}
\definecolor{purple}{HTML}{880088}

\maketitle

\begin{abstract}

We investigate the parameterized complexity of several problems formalizing cluster identification in graphs. In other words we ask whether a graph contains a large enough and sufficiently connected subgraph. We study here three relaxations of \clique{}: \sclub{} and \sclique{}, in which the relaxation is focused on the distances in respectively the cluster and the original graph, and \gcs{} in which the relaxation is made on the minimal degree in the cluster. As these three problems are known to be {\sf NP}-hard, we study here their parameterized complexities. We prove that \sclub{} and \sclique{} are {\sf NP}-hard even restricted to graphs of degeneracy $\le 3$ whenever $s \ge 3$, and to graphs of degeneracy $\le 2$ whenever $s \ge 5$, which is a strictly stronger result than its {\sf W[1]}-hardness parameterized by the degeneracy. We also obtain that these problems are solvable in polynomial time on graphs of degeneracy $1$. Concerning \gcs{}, we prove that it is {\sf W}[1]-hard parameterized by both the degeneracy, which implies the {\sf W[1]}-hardness parameterized by the number of vertices in the $\gamma$-complete-subgraph, and the number of elements outside the $\gamma$-complete subgraph.
    
\end{abstract}

\section{Introduction} 

Finding large clusters in graphs is a fundamental problem in graph theory. Among the applications, one could think of friends recommendation in social networks as it might help to find users with similar interests, as well as the applications to biological networks, where two elements in the same clusters can reasonably be considered to share a common behavior \cite{alon2007network,balasundaram2005novel,spirin2003protein}.

A very natural way to formalize the concept of clusters could be to define them as induced cliques in the given graph. The problem \clique{} of finding a large enough clique in the input graph is of paramount importance in graph computational complexity theory, and is even one of Karp's 21 {\sf NP}-complete problem \cite{karp1972}. Not only is \clique{} {\sf NP}-complete, it is also known that the size of the largest clique is {\sf NP}-hard to approximate with a factor $O(n^{1-\varepsilon})$ with $n$ the number of vertices and $\varepsilon>0$ is fixed \cite{zuckerman2006linear}.

However, asking for a clique is often too restrictive in practice, as it is reasonable to consider a subgraph to be a cluster even if a few edges are missing between some pairs of elements. Therefore, relaxations of \clique{} are considered: one can think of a relaxation by degree, connectivity, density, or distance \cite{komusiewicz2016multivariate}.

Unsurprisingly, each of these relaxations has quickly been proven to be {\sf NP}-hard \cite{bourjolly2002exact,lewis1980node,matsuda1999classifying,pattillo2013maximum,pei2005mining,veremyev2014finding}. In order to get a better understanding of these problems, it is then natural to tackle them under the paradigm of parameterized complexity. Among the relevant parameters to consider, one could think of the number of vertices of the sought cluster $k$, the number of vertices out of the cluster $\ell=n-k$, the degree of the graph $\Delta$, the $h$-index of the graph $h$, and its degeneracy $d$.

The case of \clique{} parameterized by $k$ has been treated successfully thanks to powerful meta-theorems \cite{khot2002parameterized}, which leads to its {\sf W[1]}-hardness. Concerning the parameterization by $d$, it is sufficient to observe that a clique is always contained in the neighborhood of each of its vertices to get an {\sf FPT} algorithm \cite{komusiewicz2016multivariate}. As an immediate consequence, we get that \clique{} is also {\sf FPT} when parameterized by $\Delta$ and $h$, since $d\le h\le \Delta$.
The case of $\ell$ can be treated with the very well known {\sf FPT}-reduction from {\sc Vertex-Cover} parameterized by $k$ which leads to an {\sf FPT} algorithm running in time $O(1.28^{\ell}+n^2)$ with $n$ the number of vertices, making use of the $O(1.28^k+n^2)$ algorithm for {\sc Vertex-Cover} \cite{chen2010improved}.

Unfortunately, treating the relaxations of \clique{} is usually a lot more difficult. A major technical obstruction to the study of the parameterized complexity of the relaxation of \clique{} is often that the class of cluster defined in these problems are not hereditary, i.e. stable by any vertex deletion. Therefore, the most commonly useful meta-theorems \cite{khot2002parameterized} cannot be applied, and both algorithms and reductions become more difficult to find. Despite this difficulty, the parameterized complexity of the {\sc $\gamma$-Quasi-Clique} problem, (with $\gamma \in\ ]0,1[$ a rational) -that looks for a subgraph of size at least $k$ and of density at least $\gamma$- has been completely classified regarding the parameters $k,\ell,\Delta,h$ and $d$ evoked earlier \cite{komusiewicz2015algorithmic}. This success can be attributed to the property of quasi-hereditary of the class of $\gamma$-quasi-clique (one vertex can always be removed from a $\gamma$-quasi-clique of size $>2$ to get a smaller $\gamma$-quasi-clique).

Two different distance based relaxations of \clique{}, have been introduced so far: given an integer $s\ge 2$, they both ask for a subgraph of size $k$ where every pair of vertices is at distance at most $s$. If the distance considered is the distance in the induced subgraph chosen, the problem defined thereby is known as the {\sc $s$-Club} problem, whose multivariate complexity analysis is now completed for $s=2$ and almost completed for other values of $s$ \cite{bourjolly2002exact,hartung2015structural,komusiewicz2016multivariate,schafer2012parameterized} (despite not being quasi-hereditary). If the distance is taken in the original graph instead, the problem is called the {\sc $s$-Clique} problem, which received significantly less attention than {\sc $s$-Club} (despite being hereditary). Surprisingly enough, {\sc $s$-Club} and \sclique{} with $s\ge 2$ are {\sf FPT} parameterized by $k$ (and $\ell$) \cite{komusiewicz2016multivariate,schafer2012parameterized}. This differs from the {\sc $1$-Club} =  {\sc $1$-Clique} = \clique{} problem which is known to be {\sf W[1]}-hard parameterized by $k$. Some additional results are known in the particular case of {\sc $2$-Club}: it is {\sf W[1]}-hard when parameterized by the $h$-index, and it is {\sf NP}-hard even for graphs of degeneracy at most $6$ \cite{hartung2015structural}. However, it was still open to determine if $s$-\club{} for $s\ge 3$ shares this common behavior with $2$-\club{}.

Our first contribution is to complete the study of $s$-\club{} parameterized by the degeneracy $d$. We prove in Section~\ref{sec:sclique} that, for $s\ge 3$, it is {\sf NP}-hard even on graphs of degeneracy $\le 3$. This implies that $s$-\club{} is para-{\sf NP}-hard when parameterized by $d$. This result does not only prove the {\sf W[1]}-hardness of {\sc $s$-Club} when parameterized by $d$, but also rule out the existence of any {\sf XP} algorithm (i.e. algorithm running in time $O(F(d)\times n^{F(d)})$ with $n$ the size of the graph and $F$ a computable function), under the assumption that {\sf P$\neq$NP}. Moreover, our reduction leads to a bipartite graph if $s$ is odd, and to a graph of degeneracy $2$ if $s\ge 5$, leading to the {\sf NP}-hardness on these respective classes of graphs.

In addition, we show that our reduction to $s$-\club{} with $s\ge 3$ can also be used to derive the exact same complexity results for \sclique{}. We also remark that {\sc $s$-Club} and {\sc $s$-Clique} can be solved in polynomial time on graphs of degeneracy $1$, proving that the {\sf NP}-hardness on graphs of degeneracy $2$ obtained when $s\ge 5$ is essentially optimal.

Concerning the degree based relaxations of \clique{}, the \gcs{} problem (with $\gamma \in\ ]0,1[$ a rational), where one looks for a subgraph of size has least $k$ where every vertex has a proportion at least $\gamma$ of neighbors, has received very little attention, possibly because the class of $\gamma$-complete-graphs is not even quasi-hereditary. So far, it is only known that \gcs{} is {\sf FPT} when parameterized by the $h$-index (which leads to it being {\sf FPT} when parameterized by $\Delta$) and {\sf W[1]}-hard when parameterized by $k$, but only in the case where $\gamma \geq 0.5$ \cite{baril2021hardness}.

Our second and third contributions in this paper consist in the completion of the multivariate complexity analysis of \gcs{}: we first prove in Section~\ref{sec:gcs_degeneracy} that it is {\sf W[1]}-hard when parameterized by $d$, and as an immediate corollary, we obtain that \gcs{} is also {\sf W[1]}-hard when parameterized by $k$, extending the result of Baril {\em et al.} \cite{baril2021hardness} from rational $\gamma \in [0.5,1[$ to any rational $\gamma\in\ ]0,1[$ (note that the case $\gamma=0$ corresponds to a trivial problem, and that the case $\gamma$=1 corresponds exactly to the \clique{} problem and is thus already known). Second, in Section~\ref{sec:gcs_l}, we prove that \gcs{} is {\sf W[1]}-hard when parameterized by $\ell$.
Note that trivial {\sf XP} algorithms exits for both of these parameterizations and thus the {\sf W[1]}-hardness results obtained cannot be extend to para-{\sf NP}-hardness results.

\section{Preliminaries}

The notations $\mathbb{N}$ and $\mathbb{N}^*$ design respectively the set of positive and strictly positive intergers. Given a set $V$, we denote by $\binom{V}{2}$ the set of pairs of $V$. Formally, we have $\displaystyle{\binom{V}{2}:= \{\{u,v\} \mid (u,v)\in V^2, u\neq v\}}$.

\begin{definition}

A graph $G$ is a pair $(V_G,E_G)$ where $V_G$ is a finite subset (called the set of vertices of $G$) and $E_G\subseteq \binom{V_G}{2}$ (called the set of edges of $G$). For a graph $G$, denote $\overline{E_G} = \binom{V_G}{2}\setminus E_G$.

\end{definition}

In any context where a graph is called $G$, we will use the notations $n:=|V_G|$, $m:=|E_G|$ and $\overline{m}:=|\overline{E_G}| = \binom{n}{2}-m$. Let $G=(V_G,E_G)$ be a graph for the rest of this section.

\begin{notation}

For $(u,v) \in (V_G)^2$, denote by $\deg_G(u)$ the degree of $u$ in $G$, and $\dist_G(u,v)$ the distance between $u$ and $v$ in $G$. For $r\ge 1$ and $u\in V_G$, denote by $\displaystyle{\overline{B}(u,r):=\{v\in V_G \mid \dist_G(u,v)\le r\}}$ the ball of center $u$ and radius $r$.

\end{notation}

For $S\subseteq V_G$, denote by $G[S]$ the graph induced on $G$ by $S$, i.e. the graph $(S,E_G\cap \binom{S}{2})$. Also, for any $u\in V_{G}$ (respectively $(u,v)\in (V_G)^2$), we denote by $\deg_S(u)$ (respectively $\dist_S(u,v)$) the number of neighbors of $u$ in $S$, i.e. degree of $u$ in the graph $G[S \cup \{u\}]$ (respectively the distance between $u$ and $v$ in $G[S\cup\{u,v\}]$). For $w\in V_G$, denote by $G-w$ the graph $G[V_G\setminus\{w\}]$: this is the graph obtained from $G$ after the deletion of $w$.

\begin{definition}

The {\it diameter} of a connected graph $G$ denoted by $\diam(G)$ is the maximal distance between two vertices in $G$.

\end{definition}

By an easy induction on $m$, we obtain Lemma~\ref{lem:sum_degree} (also known as the "handshaking lemma").

\begin{lem}\label{lem:sum_degree}

For any graph $G$, we have $\sum\limits_{u\in V_G} \deg_G(u) = 2m$.

\end{lem}

\subsection{Parameterized complexity}

Given an algorithmic problem $\Pi$ on a language $\Sigma^*$, a {\it parameter} of $\Pi$ is a function $\lambda:\Sigma^*\mapsto \mathbb{N}$ computable in polynomial time (where $\mathbb{N}$ is the set of positive integers). A parameterized problem is a couple of the form $(\Pi,\lambda)$ where $\Pi$ is an algorithmic problem, and $\lambda$ is a parameter of $\Pi$.

We say that an algorithm is {\sf FPT} (Fixed Parameter Tractable) parameterized by a parameter $\lambda$ if it runs in time $O( F(\lambda(x))\times \|x\|^{O(1)})$ on any instance $x$ of size $\|x\|$. Here, $F$ can be any computable function (that can be assumed to be monotonous). A parameterized problem $(\Pi,\lambda)$ is said to be {\sf FPT} if it can be solved by an {\sf FPT} algorithm. Note that a polynomial time is in particular an {\sf FPT} time for any parameter.

A {\it {\sf FPT}-reduction} from a parameterized problem $(\Pi_1,\lambda_1)$ to a parameterized problem $(\Pi_2,\lambda_2)$ is a function $R$ that maps any instance $x$ of $\Pi_1$ to an instance $R(x)$ of $\Pi_2$ that satisfies the following properties:

\begin{itemize}

    \item For all instance $x$ of $\Pi_1$, $x\in\Pi_1 \iff R(x)\in\Pi_2$.

    \item $R$ is computable in {\sf FPT} time parameterized by $\lambda_1$.

    \item For all instance $x$ of $\Pi_1$, $\lambda_2(R(x)) \le G(\lambda_1(x))$ (here, $G$ can be any computable function).

\end{itemize}

The interest of this definition is that if there exists a {\sf FPT}-reduction from $(\Pi_1,\lambda_1)$ to $(\Pi_2,\lambda_2)$ and if $(\Pi_2,\lambda_2)$ is {\sf FPT}, then so is $(\Pi_1,\lambda_1)$.

In this paper, we say that a parameterized problem $(\Pi,\lambda)$ is {\sf W[1]}-hard if it can be {\sf FPT}-reduced from (\clique{}, $k$) (where $k$ is the size of the demanded clique). Note that it is not the rigorous definition of the {\sf W[1]}-hardness in the literature, but this definition is equivalent, and sufficiently precise for the purpose of this paper.

\begin{rem}\label{rem:W1-hard_comparaison}

Take two parameters $\lambda_1$ and $\lambda_2$ on the same non-trivial (i.e., neither always yes not always no) algorithmic problem $\Pi$, such that $(\Pi,\lambda_1)$ is {\sf W[1]}-hard, and $\lambda_2\le \lambda_1$. Then, $(\Pi,\lambda_2)$ is also {\sf W[1]}-hard. More generally, if we only assume that there exists a poly-time computable function $G$ such that $\Pi$ is solvable in polynomial time on any instance $x$ that does not satisfy $\lambda_2(x) \le G(\lambda_1(x))$, then $(\Pi,\lambda_2)$ is also {\sf W[1]}-hard.

\end{rem}

\begin{proof}

In the case where $\lambda_2\le \lambda_1$, the reduction that maps any instance to itself is a {\sf FPT}-reduction.

In the more general case, the reduction $R$ consists on an instance $x$, to test in polynomial time if  $\lambda_2(x) \le G(\lambda_1(x))$ holds. In case it does not, the problem is solvable in polynomial time: solve it and return any equivalent small instance (of size $O(1)$), and it case it does, set $R(x)=x$, we indeed have $\lambda_2(R(x))\le G(\lambda_1(x))$.

\end{proof}

Analogously to the {\sf P$\neq$NP} conjecture in polynomial complexity, the central conjecture in the field of parameterized complexity is that {\sf FPT$\neq$W[1]} (the definition of the class {\sf W[1]} is omitted here). This conjecture is widely believed to be true, and implies that any {\sf W[1]}-hard problem can not be {\sf FPT} (similarly to how under {\sf P$\neq$NP}, any {\sf NP}-hard problem can not be polynomial).

A parameterized problem $(\Pi,\lambda)$ is said to be {\sf XP} if it can be solved in polynomial time on any class of instance of bounded parameter, or equivalently, if it can be solved in time $O(F(\lambda(x))\times \|x\|^{F(\lambda(x))})$ on any instance $x$ of size $\|x\|$ (here, $F$ can be any computable function). Notice that {\sf FPT} $\subseteq$ {\sf XP}, but the reverse is false under {\sf FPT}$\neq${\sf W[1]}. For instance, (\clique{}, $k$) is {\sf W[1]}-hard, but belongs to {\sf XP}, since testing every set of vertices of size $k$  of the input graph and checking if it is a clique can be done in time $O(k^2\times\binom{n}{k})=O( k^2 \times n^k)$.

\begin{rem}\label{rem:XP_comparaison}
Take two parameters $\lambda_1$ and $\lambda_2$ on the same algorithmic problem $\Pi$ such that $(\Pi,\lambda_2)$ is {\sf XP}, and $\lambda_2\le \lambda_1$. Then, $(\Pi,\lambda_1)$ is also {\sf XP}. More generally, if we only assume that there exists a poly-time computable function $G$ such that $\Pi$ is solvable in polynomial time on any instance $x$ that does not satisfy $\lambda_2(x) \le G(\lambda_1(x))$, then $(\Pi,\lambda_1)$ is also {\sf XP}.

\end{rem}

\begin{proof}

In the case where $\lambda_2\le \lambda_1$ any algorithm that is {\sf XP} parameterized by $\lambda_2$ is also {\sf XP} parameterized by $\lambda_1$.

In the more general case, on an instance $x$, one can test in polynomial time if  $\lambda_2(x) \le G(\lambda_1(x))$ holds. In case it does not, the problem is solvable in polynomial time, and in case it does, $(\Pi,\lambda_2)$ being {\sf XP} leads to an algorithm running in time $O(F(\lambda_2(x))\times \|x\|^{F(\lambda_2(x))})=O( F(G(\lambda_1(x)))\times \|x\|^{F(\lambda_2(x))})$.

\end{proof}

A parameterized problem $(\Pi,\lambda)$ is said to be para-{\sf NP}-hard if it is {\sf NP}-complete on a class of instances of $\Pi$ bounded for $\lambda$. Note that under {\sf P$\neq$NP}, a para-{\sf NP}-hard problem can not be in {\sf XP}. Note also that any para-{\sf NP}-hard problem is automatically {\sf W[1]}-hard (the reduction ensuring the para-{\sf NP}-hardness is indeed also an {\sf FPT}-reduction), even though the reverse is not true. For instance, (\clique{}, $k$) is {\sf W[1]}-hard, but is not para-{\sf NP}-hard, since it is in {\sf XP}.

\subsection{Problems studied}

To prove the {\sf W[1]}-hardness and para-{\sf NP}-hardness of relaxations of \clique{}, it seems natural to start our reduction from the \clique{} problem, which is known to be {\sf W[1]}-hard when parameterized by $k$, the size of the demanded clique.

\begin{pb}[{\clique}]

\hfill

\textbf{Input:} A graph $G$ and an integer $k$.

\textbf{Question:} Does there exist a subset $K\subseteq V_G$ with $|K|= k$, and such that $\displaystyle{\forall (u,v)\in K^2, u\neq v\implies \{u,v\}\in E_G}$ ?

\end{pb}

This problem is known to be {\sf W[1]}-hard parameterized by $k$. However, for some technical reasons, it might be useful to assume additional properties on $k$. In fact, one can notice that for any subset $I=\{ar+b\mid r \in \mathbb{N}, r\ge r_0\}$ with three constants $a,b$ and $r_0$ with $a\in \mathbb{N}^*$ and $(b,r_0)\in\mathbb{N}^2$, the problem:

\begin{pb}[{\clique}($I$)]

\hfill

\textbf{Input:} A graph $G$ and an integer $k\in I$.

\textbf{Question:} Does there exist a subset $K\subseteq V_G$ with $|K|= k$, and such that $\displaystyle{\forall (u,v)\in K^2, u\neq v\implies \{u,v\}\in E_G}$ ?

\end{pb}

is still {\sf W[1]}-hard when parameterized by $k$. Indeed, any instance $(G,k)$ of \clique{} is equivalent to the instance $(G+q,k+q)$ of \clique{} for any $q\geq 0$ (denoting by $G+q$ the graph $G$ where $q$ universal vertices have been added). By choosing $q$ such that $k+q\in I$, we can always reach an instance of \clique{}($I$). Notice also that the number $q$ of vertices to add depend only on $k$ and not on the graph $G$, which proves that \clique{}($I$) is also {\sf W[1]}-hard.

\medbreak

In the first part of this paper, we will study for $s\ge 2$ the \sclub{} and \sclique{} problem, formally defined as:

\begin{pb}[\sclub]

\hfill

\textbf{Input:} A graph $G$ and an integer $k$.

\textbf{Question:} Does there exist a subset $S\subseteq V_G$ with $|S|\ge k$, and such that $\displaystyle{\forall (u,v)\in S^2, \dist_S(u,v)\le s}$ ?
 
\end{pb}


\begin{pb}[$s$-\clique{}]

\hfill

\textbf{Input:} A graph $G$ and an integer $k$.

\textbf{Question:} Does there exist a subset $S\subseteq V_G$ with $|S|\ge k$, and such that $\displaystyle{\forall (u,v)\in S^2, \dist_G(u,v)\le s}$ ?
 
\end{pb}

Note that the only difference between \sclub{} and \sclique{} is that the distance is taken in $G[S]$ in the case of \sclub, and in $G$ in the case of \sclique{}.

A subset of vertices $S\subseteq V_G$ satisfying $\forall (u,v)\in S^2, \dist_S(u,v)\le s$ (respectively $\dist_G(u,v)\le s$) is called a {\it $s$-club of $G$} (respectively a {\it $s$-clique of $G$}). Notice that for $s=1$, the three problems \sclub{}, \sclique{}, and \clique{} are the same.

Then, given a rational $\gamma\in \ ]0,1[$, we will focus on the \gcs{} problem, formally defined as:

\begin{pb}[{\gcs}]

\hfill

\textbf{Input:} A graph $G$ and an integer $k$.

\textbf{Question:} Does there exist a subset $S\subseteq V_G$ with $|S|\ge k$, and such that $\displaystyle{\forall u\in S, \deg_S(u)\ge \gamma(|S|-1)}$ ?

\end{pb}

A subset of vertices $S\subseteq V_G$ satisfying $\forall u\in S, \deg_S(u)\ge \gamma(|S|-1)$ is called a {\it $\gamma$-complete subgraph of $G$}. The \gcs{} problem is said to be a relative relaxation of clique by degree. Note that indeed, if $\gamma=1$, \gcs{} corresponds to \clique{}.

Note that when studying the multivariate complexity of \sclub{} or \sclique{} (respectively \gcs{}), the value of $s$ (respectively $\gamma$) will always be considered to be a constant and will never be considered to be a part of the input. Thus, in the {\sf FPT}-reduction we will perform, the size of the outputs and the values of the considered parameters are allowed to depend on $s$ (respectively $\gamma$), but we won't be able to choose a particular value of $s$ (respectively $\gamma$) depending on the input.

Note also that not authorizing $s$ (respectively $\gamma$) to be part of the input strengthen our results of difficulty, since it proves that the problem with $s$ (respectively $\gamma$) as part of the input stays hard even if a particular value of $s$ (respectively $\gamma$) is fixed.

\subsection{Notations for parameters}

In this subsection, we introduce the notations and the definitions of the parameter used to study the parameterized complexities of \sclub{}, \sclique{} and \gcs{}.

\begin{itemize}

\item Consistently with the notation used in the definition of \clique{}, \sclub{}, \sclique{} and \gcs{}, the parameter denoted by $k$ will always refer to the (minimal) size of an aimed subgraph (either a clique, an $s$-club, an $s$-clique or a $\gamma$-complete-subgraph).

\item The parameter $\ell:=n-k=|V_G|-k$ designs the (maximal) number of vertices outside the aimed subgraph.

\item The parameter $\Delta:=\Delta(G)$ is the maximal degree in the input graph.

\item The parameter $h:=h(G)$ is the {\it $h$-index} of the input graph $G$, i.e. the largest integer $h$ such that $G$ has at least $h$ vertices of degree at least $h$.

\item Finally, the parameter $d:=d(G)$ designs the degeneracy of the input graph $G$, i.e. the largest integer $d$ such that every subgraph of $G$ has a vertex of degree at most $d$. A graph $G$ is $d$-degenerated (i.e. its degeneracy is upper-bounded by $d$) if, and only if, there is an order over the vertices of $G$ in which each vertex has at most $d$ inferior neighbors. Such an order is called a {\it $d$-elimination order}, or simply an elimination order. Note that an order $\leq$ over the vertices of $G$ is a $d$-elimination order if, and only if, when eliminating sequentially the vertex by decreasing order relatively to $\leq$, we eliminate a vertex of degree at most $d$ at each step.

\end{itemize}

One can remark that on any graph $G$, we have $d\le h\le \Delta$. As a consequence, every {\sf W[1]}-problem parameterized by $\Delta$ is also {\sf W[1]}-hard parameterized by both $h$ and $d$, and every {\sf FPT} problem parameterized by $d$ is automatically {\sf FPT} when parameterized by both $h$ and $\Delta$.

\subsection{Contributions}

Let $s\ge 2$ an integer and $\gamma \in\ ]0,1[$ a rational. The table in Figure~\ref{fig:state_of_the_art} sums up the state of the art on the parameterized complexities of \clique{}, \sclub{}, \sclique{} and \gcs{} parameterized by $k$, $\ell$, $\Delta$, $h$ and $d$. We omit the column representing $\Delta$ since each of these problems is {\sf FPT} when parameterized by $\Delta$ \cite{komusiewicz2016multivariate}.

\begin{figure}
\begin{tabular}{c c c c c}
    Problem & $k$ & $\ell$ & $h$ & $d$ \\ \hline
    \clique{} & {\sf W[1]}-h\cite{downey2013} & {\sf FPT}~\cite{komusiewicz2016multivariate}& {\sf FPT}~\cite{komusiewicz2016multivariate}& {\sf FPT}~\cite{komusiewicz2016multivariate}\\
    $2$-\club{} & {\sf FPT}~\cite{komusiewicz2016multivariate} & {\sf FPT}~\cite{komusiewicz2016multivariate} & {\sf W[1]}-h~\cite{hartung2015structural} & {\sf NP}-h for $d=6$~\cite{hartung2015structural} \\
    $s$-\club{} with $s\ge 3$ & {\sf FPT}~\cite{komusiewicz2016multivariate} & {\sf FPT}~\cite{komusiewicz2016multivariate}& ? & ? \\
    $s$-\clique{} & {\sf FPT}~\cite{komusiewicz2016multivariate} & {\sf FPT}~\cite{komusiewicz2016multivariate}& ? & ? \\
    \gcs{} & {\sf W[1]}-h \text{ if }$\gamma\ge \dfrac{1}{2}$\cite{baril2021hardness} & ? & {\sf FPT}~\cite{baril2021hardness} & ?  \\
\end{tabular}

\caption{Parameterized complexities of relaxations of \clique{}.}
\label{fig:state_of_the_art}
\end{figure}

\hfill

In this paper, we give five contributions:

\begin{itemize}

    \item We prove in Section~\ref{sec:sclique} that if $s\ge 3$, $s$-\club{} is {\sf NP}-hard even on graphs of degeneracy at most $3$. This implies in particular that for $s\ge 3$, $s$-\club{} is {\sf W[1]}-hard when parameterized by $d$. Moreover, if $s$ is odd, we obtain that $s$-\club{} is {\sf NP}-hard even on bipartite graphs of degeneracy at most $3$, and if $s\ge 5$, $s$-\club{} is {\sf NP}-hard even on graphs of degeneracy at most $2$. For $s\ge 5$ odd, we obtain that $s$-\club{} is {\sf NP}-hard even on bipartite graphs of degeneracy at most $2$.
    
    \item Also in Section~\ref{sec:sclique}, we prove that the results listed in the previous point also hold for \sclique{}.

    \item We prove in Section~\ref{sec:gcs_degeneracy} that \gcs{} is {\sf W[1]}-hard when parameterized by the degeneracy $d$ of the input graph.

    \item As an immediate corollary, we obtain that \gcs{} is {\sf W[1]}-hard when parameterized by $k$ regardless of the value of $\gamma$, extending the result by Baril {\em et al.} \cite{baril2021hardness} even for $\gamma < \frac{1}{2}$.

    \item Finally, we prove in Section~\ref{sec:gcs_l} that \gcs{} is {\sf W[1]}-hard when parameterized by $\ell$: the minimal number of vertices not taken in the $\gamma$-complete subgraph.

\end{itemize}

\section{\texorpdfstring{$s$}{}-club and \texorpdfstring{$s$}{}-clique {\sf NP}-hard for graphs of bounded degeneracy} 
\label{sec:sclique}

We first the study the \sclub{} and \sclique{} problems for $s\ge 3$. We establish that they are para-{\sf NP}-hard when parameterized by the degeneracy. More precisely, if $s\ge 3$ is odd, \sclub{} and \sclique{} are {\sf NP}-hard even on bipartite graphs of degeneracy $3$ (even of degeneracy $2$ if $s\ge 5$), and if $s\ge 4$ is even, \sclub{} and \sclique{} are {\sf NP}-hard even on graphs of degeneracy $3$ (even of degeneracy $2$ if $s\ge 6$).

When performing our reductions, we will need to distinguish whether $s$ is odd or even. We prove four distinct theorems in order to treat these two cases separately  (and depending on whether we study \sclub{} or \sclique{}) even though the conclusion is essentially the same (despite the graph obtained for $s$ even not being bipartite). We treat the odd case in Theorem~\ref{thm:sclub_s_odd} for \sclub{} and in Theorem~\ref{thm:sclique_s_odd} for \sclique{}, and the even case in Theorem~\ref{thm:sclub_s_even} for \sclub{} and in Theorem~\ref{thm:sclique_s_even} for \sclique{}.


\begin{thm}\label{thm:sclub_s_odd}
    For any $s \ge 3$ odd, \sclub{} is {\sf NP}-hard, even on bipartite $3$-degenerate graphs. Moreover, if $s\ge 5$, \sclub{} is {\sf NP}-hard, even on bipartite $2$-degenerate graphs.
\end{thm}

\begin{proof}

Let $s \ge 3$ be an odd integer.

We provide a {\sf NP}-hardness reduction from \clique. Let $(G,k)$ be an instance of \clique. Without loss of generality, we assume that $G$ has no isolated vertices. We build an instance $(G', k')$ of \sclub{} as in Figure~\ref{fig:2_degen_s} and Figure~\ref{fig:const_odd_sclique}.

\hfill

\underline{Construction of $G'$:}

\hfill

First, in order to decrease the degeneracy of the input graph and to increase the distances between the original vertices we subdivide $s-2$ times each edge by replacing them with a path of red vertices of length $s-1$ (i.e. with $s-1$ edges) as done in Figure~\ref{fig:2_degen_s}.

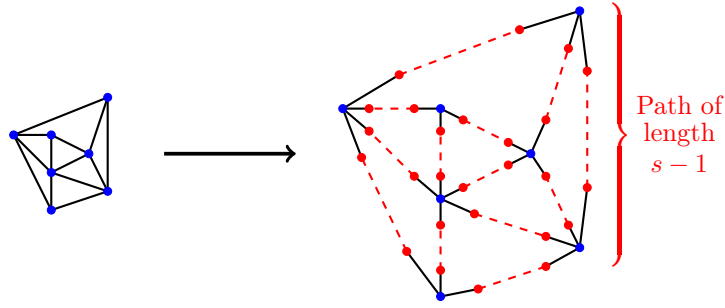
\begin{figure}[!ht]
    \centering
    \begin{tikzpicture}[scale=0.5]
    \draw[thick,black] (-0.5,-1.5)--(-0.5,-0.5);
    \draw[thick,black] (-0.5,-1.5)--(1.0,-1.0);
    \draw[thick,black] (-0.5,-1.5)--(-1.5,0.5);
    \draw[thick,black] (-0.5,-0.5)--(-0.5,0.5);
    \draw[thick,black] (-0.5,-0.5)--(1.0,-1.0);
    \draw[thick,black] (-0.5,-0.5)--(0.5,0.0);
    \draw[thick,black] (-0.5,-0.5)--(-1.5,0.5);
    \draw[thick,black] (-0.5,0.5)--(0.5,0.0);
    \draw[thick,black] (-0.5,0.5)--(-1.5,0.5);
    \draw[thick,black] (1.0,-1.0)--(0.5,0.0);
    \draw[thick,black] (1.0,-1.0)--(1.0,1.5);
    \draw[thick,black] (0.5,0.0)--(1.0,1.5);
    \draw[thick,black] (-1.5,0.5)--(1.0,1.5);
    
    \filldraw[blue] (-0.5,-1.5) circle (3 pt);
    \filldraw[blue] (-0.5,-0.5) circle (3 pt);
    \filldraw[blue] (-0.5,0.5) circle (3 pt);
    \filldraw[blue] (1.0,-1.0) circle (3 pt);
    \filldraw[blue] (0.5,0.0) circle (3 pt);
    \filldraw[blue] (-1.5,0.5) circle (3 pt);
    \filldraw[blue] (1.0,1.5) circle (3 pt);
    
    \draw[line width=0.5mm, -to] (2.5,0) -- (6,0);
    
    \node at (10.5,0) {
    \begin{tikzpicture}[scale=0.5]
    \draw[thick,black] (-1.2,-3.8)--(-1.2,-3.1);
    \draw[thick,black] (-1.2,-3.8)--(-0.2,-3.6);
    \draw[thick,black] (-1.2,-3.8)--(-2.1,-2.6);
    \draw[thick,black] (-1.2,-1.2)--(-1.2,-1.9);
    \draw[thick,black] (-1.2,-1.2)--(-1.2,-0.6);
    \draw[thick,black] (-1.2,-1.2)--(-0.3,-1.6);
    \draw[thick,black] (-1.2,-1.2)--(-0.6,-0.9);
    \draw[thick,black] (-1.2,-1.2)--(-1.9,-0.6);
    \draw[thick,black] (-1.2,1.2)--(-1.2,0.6);
    \draw[thick,black] (-1.2,1.2)--(-0.6,0.9);
    \draw[thick,black] (-1.2,1.2)--(-1.9,1.2);
    \draw[thick,black] (2.5,-2.5)--(1.6,-3.0);
    \draw[thick,black] (2.5,-2.5)--(1.6,-2.2);
    \draw[thick,black] (2.5,-2.5)--(2.2,-1.9);
    \draw[thick,black] (2.5,-2.5)--(2.7,-0.9);
    \draw[thick,black] (1.2,0.0)--(0.6,-0.3);
    \draw[thick,black] (1.2,0.0)--(0.6,0.3);
    \draw[thick,black] (1.2,0.0)--(1.6,-0.6);
    \draw[thick,black] (1.2,0.0)--(1.6,0.9);
    \draw[thick,black] (-3.8,1.2)--(-3.3,-0.1);
    \draw[thick,black] (-3.8,1.2)--(-3.1,0.6);
    \draw[thick,black] (-3.8,1.2)--(-3.1,1.2);
    \draw[thick,black] (-3.8,1.2)--(-2.3,2.1);
    \draw[thick,black] (2.5,3.8)--(2.7,2.2);
    \draw[thick,black] (2.5,3.8)--(2.2,2.8);
    \draw[thick,black] (2.5,3.8)--(0.9,3.3);
    \draw[thick,dashed,red] (-1.2,-3.1)--(-1.2,-1.9);
    \draw[thick,dashed,red] (-0.2,-3.6)--(1.6,-3.0);
    \draw[thick,dashed,red] (-2.1,-2.6)--(-3.3,-0.1);
    \draw[thick,dashed,red] (-1.2,-0.6)--(-1.2,0.6);
    \draw[thick,dashed,red] (-0.3,-1.6)--(1.6,-2.2);
    \draw[thick,dashed,red] (-0.6,-0.9)--(0.6,-0.3);
    \draw[thick,dashed,red] (-1.9,-0.6)--(-3.1,0.6);
    \draw[thick,dashed,red] (-0.6,0.9)--(0.6,0.3);
    \draw[thick,dashed,red] (-1.9,1.2)--(-3.1,1.2);
    \draw[thick,dashed,red] (2.2,-1.9)--(1.6,-0.6);
    \draw[thick,dashed,red] (2.7,-0.9)--(2.7,2.2);
    \draw[thick,dashed,red] (1.6,0.9)--(2.2,2.8);
    \draw[thick,dashed,red] (-2.3,2.1)--(0.9,3.3);
    
    \filldraw[blue] (-1.2,-3.8) circle (3 pt);
    \filldraw[blue] (-1.2,-1.2) circle (3 pt);
    \filldraw[blue] (-1.2,1.2) circle (3 pt);
    \filldraw[blue] (2.5,-2.5) circle (3 pt);
    \filldraw[blue] (1.2,0.0) circle (3 pt);
    \filldraw[blue] (-3.8,1.2) circle (3 pt);
    \filldraw[blue] (2.5,3.8) circle (3 pt);
    \filldraw[red] (-1.2,-3.1) circle (3 pt);
    \filldraw[red] (-1.2,-1.9) circle (3 pt);
    \filldraw[red] (-0.2,-3.6) circle (3 pt);
    \filldraw[red] (1.6,-3.0) circle (3 pt);
    \filldraw[red] (-2.1,-2.6) circle (3 pt);
    \filldraw[red] (-3.3,-0.1) circle (3 pt);
    \filldraw[red] (-1.2,-0.6) circle (3 pt);
    \filldraw[red] (-1.2,0.6) circle (3 pt);
    \filldraw[red] (-0.3,-1.6) circle (3 pt);
    \filldraw[red] (1.6,-2.2) circle (3 pt);
    \filldraw[red] (-0.6,-0.9) circle (3 pt);
    \filldraw[red] (0.6,-0.3) circle (3 pt);
    \filldraw[red] (-1.9,-0.6) circle (3 pt);
    \filldraw[red] (-3.1,0.6) circle (3 pt);
    \filldraw[red] (-0.6,0.9) circle (3 pt);
    \filldraw[red] (0.6,0.3) circle (3 pt);
    \filldraw[red] (-1.9,1.2) circle (3 pt);
    \filldraw[red] (-3.1,1.2) circle (3 pt);
    \filldraw[red] (2.2,-1.9) circle (3 pt);
    \filldraw[red] (1.6,-0.6) circle (3 pt);
    \filldraw[red] (2.7,-0.9) circle (3 pt);
    \filldraw[red] (2.7,2.2) circle (3 pt);
    \filldraw[red] (1.6,0.9) circle (3 pt);
    \filldraw[red] (2.2,2.8) circle (3 pt);
    \filldraw[red] (-2.3,2.1) circle (3 pt);
    \filldraw[red] (0.9,3.3) circle (3 pt);
    \end{tikzpicture}
    };

    \node[red,rotate=90] (nb_rouges) at 
    (14.5,0.5) {$\underbrace{\hspace{3.5cm}}^{}$};
    \node[red] (txt1) at 
    (16.2,1.3) {Path of};
    \node[red] (txt2) at 
    (16.2,0.5) {length};
    \node[red] (txt2) at 
    (16.2,-0.3) {$s-1$};
    \end{tikzpicture}
    \caption{2-degeneracy transformation of a graph.}
    \label{fig:2_degen_s}
\end{figure}

\hfill

\noindent The rest of the construction of $G'$ is done as follows. $V_{G'}$ contains three types of vertices:

\begin{itemize}
    \item The original vertices of $V_{G}$ which we call the blue vertices.

    \item $s-2$ red vertices for each original edge $\{u,v\} \in E_{G}$ added during the subdivision of the edge. The set formed by the red vertices is denoted $V_{R}$.

    \item Finally, we add a new vertex $y$ which we call yellow.
\end{itemize}

\noindent The vertices of $V_{G'}$ are connected by the following edges:

\begin{itemize}
    \item For each original edge $\{u,v\} \in E_{G}$, the edges of the subdivision exist in $G'$.

    \item The yellow vertex $y$ is linked to the middle vertex of each red path.
\end{itemize}

\noindent The graph obtained by this construction is illustrated in Figure~\ref{fig:const_odd_sclique}. Note that the graph $G'$ can be constructed in time $n^{O(1)}$ where $n=|V_{G}|$.

\begin{figure}[!ht]
    \centering
    \begin{tikzpicture}[scale=0.5]
    \draw[thick,red] (10.0,6.9)--(8.7,6.9);
    \draw[thick,red] (2.0,6.9)--(3.3,6.9);
    \draw[thick,red] (11.0,5.2)--(10.3,4.0);
    \draw[thick,red] (7.0,-1.7)--(7.7,-0.6);
    \draw[thick,red] (5.0,-1.7)--(4.3,-0.6);
    \draw[thick,red] (1.0,5.2)--(1.7,4.0);
    \draw[thick,red,dashed] (8.7,6.9)--(7.3,6.9);
    \draw[thick,red] (7.3,6.9)--(6.0,6.9);
    \draw[thick,red] (6.0,6.9)--(4.7,6.9);
    \draw[thick,orange] (6.0,6.9)--(6.0,3.5);
    \draw[thick,red,dashed] (4.7,6.9)--(3.3,6.9);
    \draw[thick,red,dashed] (10.3,4.0)--(9.7,2.9);
    \draw[thick,red] (9.7,2.9)--(9.0,1.7);
    \draw[thick,red] (9.0,1.7)--(8.3,0.6);
    \draw[thick,orange] (9.0,1.7)--(6.0,3.5);
    \draw[thick,red,dashed] (8.3,0.6)--(7.7,-0.6);
    \draw[thick,red,dashed] (4.3,-0.6)--(3.7,0.6);
    \draw[thick,red] (3.7,0.6)--(3.0,1.7);
    \draw[thick,red] (3.0,1.7)--(2.3,2.9);
    \draw[thick,orange] (3.0,1.7)--(6.0,3.5);
    \draw[thick,red,dashed] (2.3,2.9)--(1.7,4.0);
    
    \filldraw[blue] (10.0,6.9) circle (3 pt);
    \filldraw[blue] (2.0,6.9) circle (3 pt);
    \filldraw[blue] (11.0,5.2) circle (3 pt);
    \filldraw[blue] (7.0,-1.7) circle (3 pt);
    \filldraw[blue] (5.0,-1.7) circle (3 pt);
    \filldraw[blue] (1.0,5.2) circle (3 pt);
    \filldraw[red] (8.7,6.9) circle (3 pt);
    \filldraw[red] (7.3,6.9) circle (3 pt);
    \filldraw[red] (6.0,6.9) circle (3 pt);
    \filldraw[red] (4.7,6.9) circle (3 pt);
    \filldraw[red] (3.3,6.9) circle (3 pt);
    \filldraw[red] (10.3,4.0) circle (3 pt);
    \filldraw[red] (9.7,2.9) circle (3 pt);
    \filldraw[red] (9.0,1.7) circle (3 pt);
    \filldraw[red] (8.3,0.6) circle (3 pt);
    \filldraw[red] (7.7,-0.6) circle (3 pt);
    \filldraw[red] (4.3,-0.6) circle (3 pt);
    \filldraw[red] (3.7,0.6) circle (3 pt);
    \filldraw[red] (3.0,1.7) circle (3 pt);
    \filldraw[red] (2.3,2.9) circle (3 pt);
    \filldraw[red] (1.7,4.0) circle (3 pt);
    \filldraw[orange] (6.0,3.5) circle (3 pt) node[label=right:{$y$}] {};
    
    \node[red] (nb_rouges) at 
    (6,7.5) {$\overbrace{\hspace{4cm}}^{}$};
    \node[red] (nb_rouges2) at 
    (6,8.0) {$P_s$ of length $s-1$};
    
    \end{tikzpicture}
    \caption{Construction of $G'$ on $G$ being three disjoint edges for $s$ odd.}
    \label{fig:const_odd_sclique}
\end{figure}
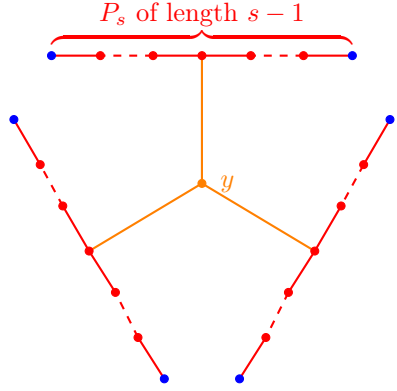

\hfill

\underline{Soundness of the reduction:}

\hfill

Let $k' := k + |V_R| + 1 = k + (s-2)m +1$. We prove that $G$ has a clique of order $k$ if, and only if, $G'$ has an $s$-club of order $k'$.

\begin{claim}\label{claim:distGodd}

For every $(u,v)\in (V_{G'})^2$ with $u\neq v$, we have:

\begin{enumerate}
    \item $\dist_{G'}(u,v) \le s$ if either $u$ or $v$ is not in $V_G$
    \item else, $\dist_{G'}(u,v) = \left\{\begin{array}{cc}
         s-1 & \text{ if }\{u,v\}\in E_G  \\
         s+1 & \text{ if }\{u,v\}\notin E_G
    \end{array} \right.$
    \item For any $(u,v)\in (V_{G'})^2$ with $\dist_{G'}(u,v)\le s$, there exists path $P$ from $u$ to $v$ of length $\le s$ such that $P\setminus\{u,v\}$ does not intersect $V_G$: ie. $\dist_{V_{G'}\setminus V_G}(u,v)\le s$.
\end{enumerate}

\end{claim}

\begin{proofclaim2}

\begin{enumerate}

\item We notice that $(V_{G'}\setminus V_G)$ is included in $\overline{B}_{G'}\left(y,\dfrac{s-1}{2}\right)$, the closed ball of center $y$ and radius $\dfrac{s-1}{2}$. Thus, every pair of non-blue vertices of $G$' (i.e. in $V_{G'}\setminus V_G$) is at distance at most $s-1$. Now, since every blue vertex $u\in V_G$ has a neighbor in $V_R$, it is at a distance lower than $s$ from any non-blue vertex.

\item By construction, for all pair of blue vertices that forms an edge in $G$: $\{u,v\}\in E_G$, we have $\dist_{G'}(u,v) = s-1$ (taking the red path of length $s-1$ that comes from the subdivision of the original edge $\{u,v\}$).

Conversely, any shortest path between two blue vertices $u$ and $v$ in $V_{G}$ such that $\{u,v\} \notin E_{G}$ is, starting from $u$, first getting to $y$ in $\dfrac{s-1}{2}+1$ steps, and symmetrically getting to $v$ in $1+\dfrac{s-1}{2}$ steps, which gives a path of length $s+1$. Indeed, not using $y$ would imply to take the red vertices that come from the subdivisions of at least two edges in $G$, which is a path of length at least $2(s-1)\ge s+1$, since $s\ge 3$. 

\item The paths given by what precedes in the cases $\dist_{G'}(u,v)\le s$ satisfy the requirement. \qedclaim

\end{enumerate}

\end{proofclaim2}

\begin{itemize}
    \item[$\implies$] Suppose $G$ has a clique $K \subseteq V_G$ of order $k$. Consider in $G'$ the set $S:=K \cup V_R \cup \{y\}$. Note that $|S|=k'$. According to Claim~\ref{claim:distGodd}.1 and \ref{claim:distGodd}.2, since $S$ does not contain any pair $\{u,v\}\notin E_G$ (because $K$ is a clique), every pair of vertices is at distance $\le s$ in $G'$. Note that since $V_R \cup \{y\}\subseteq S$, we get by Claim~\ref{claim:distGodd}.3 that every pair of vertices in $S$ is at distance $\le s$ in $G[S]$. Therefore, $S$ is an $s$-club of $G'$.

    \item[$\impliedby$] Reversely, if $G'$ contains an $s$-club $S$ of order at least $k'$: let $K:=S\cap V_G$. As $|V_{G'}\setminus V_G| = k' - k$, we have $|K| \ge k$. Using that $S$ is an $s$-club and that the distance between any two vertices in $S$ is necessarily greater than their distance in $G'$, we get by Claim~\ref{claim:distGodd}.1 and \ref{claim:distGodd}.2 that $\forall (u,v)\in K^2$ with $u\neq v$, $\{u,v\}\in E_G$, i.e. $K$ is a clique of $G$.
\end{itemize}

\hfill

\underline{Degeneracy of $G'$ and bipartition:}

\hfill

We prove that $G'$ has degeneracy at most $3$:

\begin{itemize}
    \item First, remove the red vertices in any order: they all have degree $3$ or less.
    \item Then, we are left with an independent set, remove them in an arbitrary order.
\end{itemize}

Moreover, if $s\ge 5$, by beginning with the elimination of red vertices that are non-adjacent to $y$, we start with degree $2$ vertices. Then, we can remove the red vertices adjacent to $y$ that now have degree $1$. Then, we are left with an independent set. This proves that the degeneracy of $G'$ is bounded by $2$ if $s\ge 5$.

We prove that $G'$ is bipartite by showing that it does not contain any odd cycle.
Let $C$ be a cycle of $G'$. First it is easy to see that the subgraph induced by the red and yellow vertices is a tree. Hence, $C$ contains at least a blue vertex. Second, let $P$ be a simple path (or cycle) linking two consecutive blue vertices (possibly the same) in this cycle (possibly, P=C) using thus only red and yellow vertices. Either $P$ does not use $y$: it is an edge subdivided, i.e. a path of length $s-1$. Or $P$ uses $y$ and is constituted of: half of a subdivided edge, i.e. a path of length $\dfrac{s-1}{2}$, two edges containing $y$ and another half of another subdivided edge. Thus, the total length of $P$ is $s+1$. In both cases the length of $P$ is even. Since $C$ contains a blue vertex, it is a concatenation of paths between blue vertices. Hence, the length of $C$ is a sum of even integers and thus is even.

\end{proof}

We now adapt the proof to also extend the result of Theorem~\ref{thm:sclub_s_odd} to \sclique{}, in the form of Theorem~\ref{thm:sclique_s_odd}.

\begin{thm}\label{thm:sclique_s_odd}
    For any $s \ge 3$ odd, \sclique{} is {\sf NP}-hard, even on bipartite $3$-degenerate graphs. Moreover, if $s\ge 5$, \sclique{} is {\sf NP}-hard, even on bipartite $2$-degenerate graphs.
\end{thm}

\begin{proof}
We provide a {\sf NP}-hardness reduction from \clique. Let $(G,k)$ be an instance of \clique.

Consider the graph $G'$ and the integer $k'$ defined in the proof of Theorem~\ref{thm:sclub_s_odd}. We already know that $G'$ is bipartite, and we already have studied the degeneracy of $G'$.

There only remains to prove that $G$ has a clique of size $k$ if, and only if, $G'$ has an $s$-clique of size $k'$. We will consider exactly the same sets as in the proof of Theorem~\ref{thm:sclub_s_odd}.

\begin{itemize}
    \item[$\implies$] Suppose $G$ has a clique $K \subseteq V_G$ of order $k$. Consider in $G'$ the set $S=K \cup V_R \cup \{y\}$. Note that $|S|=k'$. According to Claim~\ref{claim:distGodd}.1 and \ref{claim:distGodd}.2, since $S$ does not contain any pair $\{u,v\}\notin E_G$ (because $K$ is a clique), every pair of vertices is at distance $\le s$ in $G'$: $S$ is an $s$-clique.

    \item[$\impliedby$] Reversely, if $G'$ contains an $s$-clique $S$ of order at least $k'$: let $K=S\cap V_G$. As $|V_{G'}\setminus V_G| = k' - k$, we have $|K| \ge k$. Using that $S$ is an $s$-clique, we get by Claim~\ref{claim:distGodd}.1 and \ref{claim:distGodd}.2 that $\forall (u,v)\in K^2$ with $u\neq v$, $\{u,v\}\in E_G$, i.e. $K$ is a clique of $G$.
\end{itemize}

\end{proof}

The reduction for $s$ even is slightly more complicated, but still leads to a graph of degeneracy $\le 3$ (even of degeneracy $\le 2$ if $s\ge 6$), even though the graph obtained is not bipartite. Note that the case $s=2$ is covered neither by Theorem~\ref{thm:sclub_s_even}, nor is it by Theorem~\ref{thm:sclique_s_even}.

\begin{thm}\label{thm:sclub_s_even}
    For any $s \ge 4$ even, \sclub{} is {\sf NP}-hard even on $3$-degenerate graphs. Moreover, if $s\ge 6$, \sclub{} is {\sf NP}-hard even on $2$-degenerate graphs.
\end{thm}

\begin{proof}

Let $s \ge 4$ be an even integer.

We provide a {\sf NP}-hardness reduction from \clique. Let $(G,k)$ be an instance of \clique. Without loss of generality, we assume that $G$ has no isolated vertices and at least $2$ edges. We build an instance $(G', k')$ of \sclub{}, as in Figure~\ref{fig:2_degen_s} and Figure~\ref{fig:const_even_sclique}.

\hfill

\underline{Construction of $G'$:}

\hfill

\noindent We start again by subdividing $s-2$ times each original edge of $G$ as done in Figure~\ref{fig:2_degen_s}.

\hfill

\noindent The rest of the construction of $G'$ is done as follows. $V_{G'}$ contains four types of vertices:

\begin{itemize}
    \item The original vertices of $V_{G}$ which we call the blue vertices.

    \item $s-2$ red vertices for each original edge $(u,v) \in E_{G}$ added during the subdivision of the edge we note these vertices respectively $e_{u,v}^1,...,e_{u,v}^{s-2}$. The set formed by the red vertices is denoted $V_R$.

    \item For any pair of red vertices $\{e_{u,v}^{i}, e_{u',v'}^{j}\}$ with $1\le i,j\le s-2$ and $\{u,v\} \neq \{u',v'\}$, we add $s-2$ new vertices which we call the green vertices.

    \item Finally, we add a new yellow vertex $y$.
\end{itemize}

\noindent The vertices of $V_{G'}$ are linked by the following edges:

\begin{itemize}
    \item For each original edge $(u,v) \in E_{G}$, the edges of the subdivision exist in $G'$. i.e. $u,v$ and the $s-2$ corresponding red vertices $e_{u,v}^1,...,e_{u,v}^{s-2}$ form a path $u,e_{u,v}^1,...,e_{u,v}^{s-2},v$ of length $s-1$ between $u$ and $v$.

    \item For each pair of red vertices $\{e_{u,v}^{i}, e_{u',v'}^{j}\}$ with $1\le i,j\le s-2$ and $\{u,v\} \neq \{u',v'\}$, $e_{u,v}^{i}$, $e_{u',v'}^{j}$ and the $s-2$ corresponding green vertices form a path of length $s-1$ between $e_{u,v}^{i}$ and $e_{u',v'}^{j}$.

    \item The yellow vertex $y$ is linked to the two middle vertices of each green path.
\end{itemize}

See Figure~\ref{fig:const_even_sclique} for an illustration of the construction of $G'$. Note that $G'$ can be constructed in time $n^{O(1)}$, where $n=|V_{G}|$.

\begin{figure}
    \centering
    \begin{tikzpicture}[scale=0.5]
    \draw[thick,red] (0.0,0.0)--(2.0,0.0);
    \draw[thick,red] (10.0,0.0)--(8.0,0.0);
    \draw[thick,red] (0.0,10.0)--(2.0,10.0);
    \draw[thick,red] (10.0,10.0)--(8.0,10.0);
    \draw[thick,red] (2.0,0.0)--(4.0,0.0);
    \draw[thick,dotted,vert] (2.0,0.0)--(2.0,10.0);
    \draw[thick,dotted,vert] (2.0,0.0)--(4.0,10.0);
    \draw[thick,dotted,vert] (2.0,0.0)--(6.0,10.0);
    \draw[thick,dotted,vert] (2.0,0.0)--(8.0,10.0);
    \draw[thick,red,dashed] (4.0,0.0)--(6.0,0.0);
    \draw[thick,dotted,vert] (4.0,0.0)--(2.0,10.0);
    \draw[thick,dotted,vert] (4.0,0.0)--(4.0,10.0);
    \draw[thick,dotted,vert] (4.0,0.0)--(6.0,10.0);
    \draw[thick,dotted,vert] (4.0,0.0)--(8.0,10.0);
    \draw[thick,red] (6.0,0.0)--(8.0,0.0);
    \draw[thick,dotted,vert] (6.0,0.0)--(2.0,10.0);
    \draw[thick,dotted,vert] (6.0,0.0)--(4.0,10.0);
    \draw[thick,dotted,vert] (6.0,0.0)--(6.0,10.0);
    \draw[thick,dotted,vert] (6.0,0.0)--(8.0,10.0);
    \draw[thick,dotted,vert] (8.0,0.0)--(2.0,10.0);
    \draw[thick,dotted,vert] (8.0,0.0)--(4.0,10.0);
    \draw[thick,dotted,vert] (8.0,0.0)--(6.0,10.0);
    \draw[thick,dotted,vert] (8.0,0.0)--(8.0,10.0);
    \draw[thick,red] (2.0,10.0)--(4.0,10.0);
    \draw[thick,red,dashed] (4.0,10.0)--(6.0,10.0);
    \draw[thick,red] (6.0,10.0)--(8.0,10.0);
    
    \draw[thick,orange] (5,5)--(12.5,5.0);
    \draw[thick,orange] (8.0,4)--(12.5,5.0);
    \draw[thick,orange] (8.0,6)--(12.5,5.0);
    
    \draw[thick,vert,fill=white] (5,5) ellipse (4.2cm and 1.8cm);
    
    \draw[thick,vert] (2.0,4.2)--(2.0,5.8);
    \draw[thick,vert] (3.2,4.2)--(3.2,5.8);
    \draw[thick,vert] (4.4,4.2)--(4.4,5.8);
    \draw[thick,vert] (5.6,4.2)--(5.6,5.8);
    \draw[thick,vert] (6.8,4.2)--(6.8,5.8);
    \draw[thick,vert] (8.0,4.2)--(8.0,5.8);
    
    \filldraw[vert] (2.0,4.2) circle (3 pt);
    \filldraw[vert] (2.0,5.8) circle (3 pt);
    \filldraw[vert] (3.2,4.2) circle (3 pt);
    \filldraw[vert] (3.2,5.8) circle (3 pt);
    \filldraw[vert] (4.4,4.2) circle (3 pt);
    \filldraw[vert] (4.4,5.8) circle (3 pt);
    \filldraw[vert] (5.6,4.2) circle (3 pt);
    \filldraw[vert] (5.6,5.8) circle (3 pt);
    \filldraw[vert] (6.8,4.2) circle (3 pt);
    \filldraw[vert] (6.8,5.8) circle (3 pt);
    \filldraw[vert] (8.0,4.2) circle (3 pt);
    \filldraw[vert] (8.0,5.8) circle (3 pt);
    
    \filldraw[blue] (0.0,0.0) circle (3 pt);
    \filldraw[blue] (10.0,0.0) circle (3 pt);
    \filldraw[blue] (0.0,10.0) circle (3 pt);
    \filldraw[blue] (10.0,10.0) circle (3 pt);
    \filldraw[red] (2.0,0.0) circle (3 pt);
    \filldraw[red] (4.0,0.0) circle (3 pt);
    \filldraw[red] (6.0,0.0) circle (3 pt);
    \filldraw[red] (8.0,0.0) circle (3 pt);
    \filldraw[red] (2.0,10.0) circle (3 pt);
    \filldraw[red] (4.0,10.0) circle (3 pt);
    \filldraw[red] (6.0,10.0) circle (3 pt);
    \filldraw[red] (8.0,10.0) circle (3 pt);
    \filldraw[orange] (12.5,5.0) circle (3 pt) node[label=right:{$y$}] {};
    
    \path[every right delimiter/.style={vert}] 
    (-2,7.5) node[vert,matrix of nodes,right delimiter={\{}] (m1) {Paths of\\length\\$\displaystyle \dfrac{s-2}{2}$\\~\\};
    
    \path[every right delimiter/.style={vert}] 
    (-2,2.5) node[vert,matrix of nodes,right delimiter={\{}] (m2) {Paths of\\length\\$\displaystyle \dfrac{s-2}{2}$\\~\\};
    
    \node[red] (nb_rouges) at 
    (5,10.7) {$\overbrace{\hspace{5cm}}^{}$};
    \node[red] (nb_rouges2) at 
    (5,11.5) {$P_s$ of length $s-1$};
    
    \end{tikzpicture}
    \caption{Construction of $G'$ on $G$ being two distinct edges, for $s$ even.}
    \label{fig:const_even_sclique}
\end{figure}
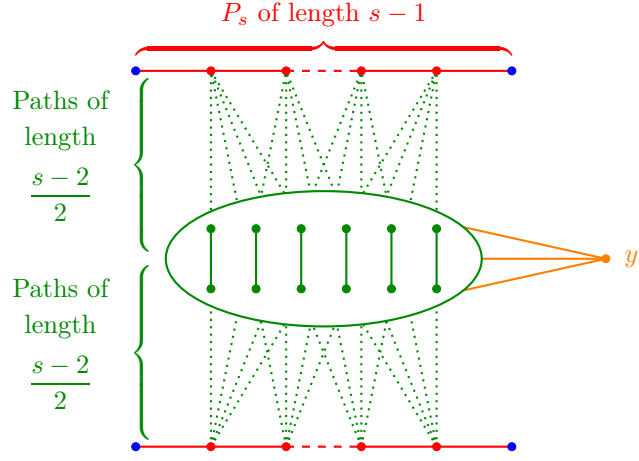

\hfill

\underline{Soundness of the reduction:}

\hfill

Let $k' := k + (s-2)m + (s-2)\binom{(s-2)m}{2} +1 $, note that there are $(s-2)m$ red vertices and $(s-2)\binom{(s-2)m}{2}$ green vertices. We prove that $G$ has a clique of order $k$ if, and only if, $G'$ has an $s$-club of order $k'$.

\begin{claim}\label{claim:distGeven}

For every $(u,v)\in (V_{G'})^2$ with $u\neq v$, we have:

\begin{enumerate}
    \item $\dist_{G'}(u,v) \le s$ if either $u$ or $v$ is not in $V_G$
    \item else, $\dist_{G'}(u,v) = \left\{\begin{array}{cc}
         s-1 & \text{ if }\{u,v\}\in E_G  \\
         s+1 & \text{ if }\{u,v\}\notin E_G 
    \end{array} \right.$
    \item For any $(u,v)\in (V_{G'})^2$ with $\dist_{G'}(u,v)\le s$, there exists a path $P$ from $u$ to $v$ of length $\le s$ such that $P\setminus\{u,v\}$ does not intersect $V_G$: ie. $\dist_{V_{G'}\setminus V_G}(u,v)\le s$.
\end{enumerate}

\end{claim}

\begin{proofclaim2}

\begin{enumerate}

\item We notice the green vertices are all contained in $\overline{B}_{G'}\left(y,\dfrac{s-2}{2}\right)$ (the closed ball of center $y$ and radius $\dfrac{s-2}{2}$). Thus, every pair of green or yellow vertices is at distance $\le s-2$. Now, since every vertex $u\in V_{G'}$ is at distance at most $2$ of a green vertex, it is at distance at most $s$ of any green or yellow vertex. Moreover, two red vertices are at distance $\le s-1$ by construction. Since any blue vertex $u\in V_G$ is adjacent to a red vertex, it is at distance at most $s$ from any red vertex. All cases have been treated.

\item By construction, for all $\{u,v\}\in E_G$, $\dist_{G'}(u,v) = s-1$ (taking the red path of length $s-1$ that comes from the subdivision of the original edge $\{u,v\}$).

Conversely, since $s>2$, any shortest path between two blue vertices $u$ and $v$ in $V_{G}$ such that $\{u,v\} \notin E_{G}$ is, starting from $u$, first to get to a red neighbor, then to go to a red neighbor of $v$ (in $s-1$ steps), and then to get to $v$. This leads to a path of length $s+1$.

\item The paths given by what precedes in the cases $\dist_{G'}(u,v)\le s$ satisfy the requirement.
\qedclaim

\end{enumerate}
\end{proofclaim2}

\begin{itemize}
    \item[$\implies$] Suppose $G$ has a clique $K \subseteq V_G$ of order $k$. Consider in $G'$ the set $S=V_{G'} \setminus (V_G \setminus K)$, i.e. all vertices except the blue ones that are not in $K$. Note that $|S|=k'$. According to Claim~\ref{claim:distGeven}.1 and \ref{claim:distGeven}.2, since $S$ does not contain any pair $\{u,v\}\notin E_G$ (because $K$ is a clique), every pair of vertices is at distance $\le s$ in $G'$. Note that since $S$ contain every non-blue vertices, we get by Claim~\ref{claim:distGeven}.3 that every pair of vertices in $S$ is at distance $\le s$ in $G[S]$: $S$ is an $s$-club of $G'$.

    \item[$\impliedby$] Reversely, if $G'$ contains an $s$-club $S$ of order at least $k'$: let $K=S\cap V_G$. As $|V_{G'}\setminus V_G| = k' - k$, we have $|K| \ge k$. Using that $S$ is an $s$-club and that the distance between any two vertices in $S$ is necessarily greater than their distance in $G'$, we get by Claim~\ref{claim:distGeven}.1 and \ref{claim:distGeven}.2 that $\forall (u,v)\in K^2$ with $u\neq v$, $\{u,v\}\in E_G$, i.e. $K$ is a clique of $G$.
\end{itemize}

\hfill

\underline{Degeneracy of $G'$:}

\hfill

We prove that $G'$ has degeneracy at most $3$:

\begin{itemize}
    \item First, remove the green vertices: they all have degree $3$ or less.
    \item Then, remove the red vertices: they now all have degree $2$.
    \item Finally, we are left with an independent set (because $s\ge 4$), remove the remaining vertices arbitrarily.
\end{itemize}

Moreover, if $s\ge 6$, one can begin with the elimination of green vertices that are not adjacent to $y$, and therefore have degree $2$. Then, we can remove the green vertices adjacent to $y$ and the red vertices that now have degree $2$. Finally, we are left with an independent set. This proves that the degeneracy of $G'$ is bounded by $2$ if $s\ge 6$.

\end{proof}

We now adapt the proof to also extend the result Theorem~\ref{thm:sclub_s_even} to \sclique{}, in the form of Theorem~\ref{thm:sclique_s_even}.

\begin{thm}\label{thm:sclique_s_even}
    For any $s \ge 4$ even, \sclique{} is {\sf NP}-hard even on $3$-degenerate graphs. Moreover, if $s\ge 6$, \sclique{} is {\sf NP}-hard even on $2$-degenerate graphs.
\end{thm}

\begin{proof}
We provide a {\sf NP}-hardness reduction from \clique. Let $(G,k)$ be an instance of \clique.

Consider the graph $G'$ and the integer $k'$ defined in the proof of Theorem~\ref{thm:sclub_s_even}. We have already studied the degeneracy of $G'$.

There only remains to prove that $G$ has a clique of size $k$ if, and only if, $G'$ has an $s$-clique of size $k'$. We will consider exactly the same sets as in the proof of Theorem~\ref{thm:sclub_s_even}.

\begin{itemize}
    \item[$\implies$] Suppose $G$ has a clique $K \subseteq V_G$ of order $k$. Consider in $G'$ the set $S=K \cup V_R \cup \{y\}$. Note that $|S|=k'$. According to Claim~\ref{claim:distGeven}.1 and \ref{claim:distGeven}.2, since $S$ does not contain any pair $\{u,v\}\notin E_G$ (because $K$ is a clique), every pair of vertices is at distance $\le s$ in $G'$: $S$ is an $s$-clique.

    \item[$\impliedby$] Reversely, if $G'$ contains an $s$-clique $S$ of order at least $k'$: let $K=S\cap V_G$. As $|V_{G'}\setminus V_G| = k' - k$, we have $|K| \ge k$. Using that $S$ is an $s$-clique, we get by Claim~\ref{claim:distGeven}.1 and \ref{claim:distGeven}.2 that $\forall (u,v)\in K^2$ with $u\neq v$, $\{u,v\}\in E_G$, i.e. $K$ is a clique of $G$.
\end{itemize}

\end{proof}

As $1$-clubs are exactly cliques, and as \clique{} is FPT parameterized by the degeneracy~\cite{Eppstein2010}, and as $2$-\club{} is known to be {\sf NP}-hard even on graphs of degeneracy $6$ \cite{schafer2012parameterized}\, Theorem~\ref{thm:sclub_s_odd} and Theorem~\ref{thm:sclub_s_even} ends the study of \sclub{} parameterized by $d$. For \sclique{}, the complexity of {\sc $2$-Clique} parameterized by the degeneracy still remains open.

\subsection{\texorpdfstring{\sclub{}}{} and \texorpdfstring{\sclique{}}{} are polynomial for \texorpdfstring{$d=1$}{}}

We prove here that for any $s\ge 1$, \sclub{} and \sclique{} is solvable in polynomial time on graphs of degeneracy $1$.
To do so, notice that graphs of degeneracy $1$ are exactly the forests, and that an $s$-club (respectively an $s$-clique) of a graph $G$ is always contained in a connected component of $G$: we can thus only treat the case where the input graph is a tree $T$. First, we notice that in the case of tree the \sclub{} and \sclique{} problems are equivalent as proven in Corollary~\ref{cor:tree_sclique_sclub}. Except for Corollary~\ref{cor:tree_sclique_sclub} and for the rest of this section we only talk about $s$-cliques but all results are also valid for $s$-clubs.

\begin{lem} \label{lem:shortest_path_sclique}
Let $T=(V_{T},E_{T})$ be a tree. For all $(u,v)\in (V_{T})^2$ and all $w$ on the unique simple path between $u$ and $v$ it holds for all $x\in V_T$ that $w$ is on the unique simple path between $u$ and $x$ or on the unique simple path between $v$ and $x$
\end{lem}

\begin{proof}
For all $u,v \in V_{T}$, we denote by $P_{u,v}$ the unique simple path between $u$ and $v$.
Let $T,u,v,w,x$ be as described in the lemma. If $w=u$ or $w=v$ the result holds. We assume that $w \notin \{u,v\}$. Since $w$ is on $P_{u,v}$, $u$ and $v$ are in two different connected component of $T-w$ (when removing $w$, we disconnect $u$ and $v$). Let's note $C_{u}$ the connected component of $T-w$ containing $u$ and $C_{v}$ the one containing $v$.

\begin{itemize}
    \item If $x \notin C_{u}$. Then, $w$ is on $P_{u,x}$ the unique simple path between $u$ and $x$ in $T$.

    \item If $x \notin C_{v}$. Then, $w$ is on $P_{v,x}$ the unique simple path between $v$ and $x$ in $T$.
\end{itemize}

\noindent Since $C_{u}$ and $C_{v}$ are disjoints $x$ cannot belong in both of them and the result holds.

\end{proof}

Corollary \ref{cor:tree_sclique_sclub} establishes the equivalence between maximal $s$-cliques and maximal $s$-sclubs in trees, justifying that we only treat the case of $s$-cliques in what will follow.

\begin{cor} \label{cor:tree_sclique_sclub}
Let $T=(V_{T},E_{T})$ be a tree, the maximal $s$-cliques of $T$ are $s$-clubs and conversely.
\end{cor}

\begin{proof}

Let $S$ be a maximal $s$-clique of $T$. Let $u,v \in S$ and let $w$ on the unique simple path between $u$ and $v$. Let's prove that $S \cup \{w\}$ is an $s$-clique of $T$. Let $x \in S$. By Lemma \ref{lem:shortest_path_sclique}, we only have two cases: 

\begin{itemize}
    \item[-] If $w$ is on the unique simple path between $u$ and $x$ then, $\dist_{T}(w,x) \leq \dist_{T}(u,x) \leq s$.

    \item[-] Otherwise, $w$ is on the unique simple path between $v$ and $x$, so $\dist_{T}(w,x) \leq \dist_{T}(u,x) \leq s$.
\end{itemize}

In both cases $\dist_{T}(w,x) \leq s$. Since the result holds for all $x \in S$, $S \cup \{w\}$ is an $s$-clique of $T$.

\hfill

By maximality of $S$, $w \in S$. $S$ necessarily contains the paths between each pair of its vertices. Hence, $S$ is an $s$-club.

\hfill

Conversely $s$-clubs are particular cases of $s$-cliques so the result holds.

\end{proof}

Now we will prove that maximal $s$-cliques of $T$ always come in three types: $T$ itself (if, and only if, $\diam(T)\le s$), a ball of diameter $\dfrac{s}{2}$ (if $s$ is even) or a union of two balls of diameter $\dfrac{s-1}{2}$ (if $s$ is odd).

\begin{claim}\label{claim:s-clique_tree}

Let $s\ge 2$ and $T$ a tree with $\diam(T) > s$, then every maximal $s$-clique of $T$ is of the form:

\begin{itemize}

    \item $\overline{B}\left(u,\dfrac{s}{2}\right)$ with $u\in V_T$ if $s$ is even.

    \item $\overline{B}\left(u,\dfrac{s-1}{2}\right) \cup \overline{B}\left(v,\dfrac{s-1}{2}\right)$ with $\{u,v\}\in E_T$ if $s$ is odd.
    
\end{itemize}

\end{claim}

\newcommand{\len}{\operatorname{len}}

\begin{proof}
Given a path $P$ in the tree $T$, we denote by $\len(P)$ its length (i.e. its number of edges). Notice that if $P$ is a path with $p\ge 1$ vertices, then $\len(P)=p-1$. Also, for $u,v \in V_{T}$ we denote by $P_{u,v}$ the unique simple path between $u$ and $v$.

\noindent First, note that for any $u\in V_T$ (respectively $\{u,v\} \in E_{T}$), if $s$ is even, $\overline{B}(u,\dfrac{s}{2})$  (respectively, if $s$ is odd, $\overline{B}(u,\dfrac{s-1}{2}) \cup \overline{B}(v,\dfrac{s-1}{2})$) is always an $s$-clique of $T$. So we only have to prove that any maximal $s$-clique is included in one of these sets and by maximality, the equality will hold.

\hfill

\noindent Take $S\subseteq V_T$ inducing a maximal $s$-clique. There exists $(u,v)\in S^2$ that are at distance exactly $s$ in $T$ because otherwise, $S$ induces a $(s-1)$-clique, and since $T$ is connected and $S\neq V_T$ (because $\diam(T)>s$), adding to $S$ any $w\in V_T\setminus S$ neighbor of $S$ would result in a strictly larger $s$-clique, contradicting the maximality of $S$.

\hfill

\noindent Let $w$ be the middle vertex of the path $P_{u,v}$ if $s$ is even and let $w,w'$ be the two middle vertices of $P_{u,v}$ if $s$ is odd ($w$ is closer to $u$ and $w'$ is closer to $v$).

\hfill

\begin{itemize}
    \item \underline{Assume that $s$ is even:}
    We will prove that $\displaystyle S \subseteq \overline{B}\left(w,\dfrac{s}{2}\right)$.
    
    Let $x\in S$. Thanks to Lemma~\ref{lem:shortest_path_sclique} it holds that either $w$ is on the path $P_{u,x}$ between $u$ and $x$ or it is on the path $P_{v,x}$ between $v$ and $x$. In the first case, it holds that $P_{u,x}$ is the concatenation of $P_{u,w}$ and $P_{w,x}$ thus,
    $$ \len(P_{w,x}) = \len(P_{u,x}) - \len(P_{u,w}) \leq s-\frac{s}{2} \leq \frac{s}{2}. $$
    
    The second case is symmetric. In both cases it holds that $\dist_{T}(w,x) \leq \frac{s}{2}$.
    
    This proves that $\displaystyle S\subseteq\overline{B}\left(w,\dfrac{s}{2}\right)$. By maximality of $S$, the equality holds.
    
    \hfill

    \item \underline{Assume that $s$ is odd:}
    We will prove that $\displaystyle S \subseteq \overline{B}\left(w,\dfrac{s-1}{2}\right) \cup \overline{B}\left(w',\dfrac{s-1}{2}\right)$.
    
    Let $x\in S$. After the concatenation of $P_{u,x}$ and $P_{x,v}$ we obtain a (not necessarily simple) path between $u$ and $v$. Since $T$ is a tree (and thus is bipartite), the length of this path has the same parity as $\len(P_{u,v})$ which is odd. Thus $\len(P_{u,x}) + \len(P_{v,x})$ is odd. Also, $\len(P_{u,x}) + \len(P_{v,x}) \leq 2s$ by definition of an $s$-clique. So $\len(P_{u,x}) + \len(P_{v,x}) \leq 2s-1$. Thus, either $\len(P_{u,x}) \leq s-1$ or $\len(P_{v,x}) \leq s-1$. Let's assume the former, the latter case is symmetrical. Using again Lemma~\ref{lem:shortest_path_sclique}, it holds that either $w$ is on the path between $u$ and $x$, in which case:
    $$ \len(P_{w,x}) = \len(P_{u,x}) - \len(P_{u,w}) \leq s-1 - \frac{s-1}{2} \leq \frac{s-1}{2}. $$
    Or it is on the path between $v$ and $x$, in which case:
    $$ \len(P_{w,x}) = \len(P_{v,x}) - \len(P_{v,w}) \leq s - \frac{s+1}{2} \leq \frac{s-1}{2}. $$
    
    Thus, $x \in \overline{B}\left(w,\dfrac{s-1}{2}\right)$. Assuming $\len(P_{v,x}) \leq s-1$ (instead of $\len(P_{u,x})\le s-1$) would have given $x \in \overline{B}\left(w',\dfrac{s-1}{2}\right)$.
    
    So, $\displaystyle S \subseteq \overline{B}\left(w,\dfrac{s-1}{2}\right) \cup \overline{B}\left(w',\dfrac{s-1}{2}\right)$. By maximality of $S$, the equality holds.
\end{itemize}
\end{proof}

Now, we propose an algorithm to solve both \sclub{} and \sclique{} on graphs of degeneracy $1$, in the form of Algorithm~\ref{algo:s-clique_degeneracy=1}.

\begin{thm}\label{thm:sclique_poly_d=1}

For $s\geq 1$, \sclub{} and \sclique{} can be solved in polynomial time on graphs of degeneracy $1$.

\end{thm}

\begin{proof}

We have by Claim~\ref{claim:s-clique_tree} that Algorithm~\ref{algo:s-clique_degeneracy=1} solves the \sclique{} problem on trees, and we have by Corollary \ref{cor:tree_sclique_sclub} that it also solves the \sclub{} problems on trees, and thus on forests since an $s$-clique (respectively an $s$-club) of a graph $G$ must be contained in a connected component of $G$. Note that forests are exactly the graphs of degeneracy $1$. 

\end{proof}

The result of Theorem~\ref{thm:sclique_poly_d=1} ensures that the {\sf NP}-hardness of \sclub{} and \sclique{} on graphs of degeneracy $2$ for $s\ge 5$ obtained in Theorem~\ref{thm:sclub_s_odd} and Theorem~\ref{thm:sclique_s_odd} and for $s\ge 6$ in Theorem~\ref{thm:sclub_s_even} and Theorem~\ref{thm:sclique_s_even} are essentially optimal, in the sense that they can not be improved to the {\sf NP}-hardness of \sclub{} and \sclique{} on graphs of degeneracy $1$, unless {\sf P=NP}.

\begin{algorithm}
\label{algo:s-clique_degeneracy=1}
\caption{Algorithm that solves both \sclub{} and \sclique{} on forests in polynomial time}

\KwData{A graph $G$ of degeneracy $1$ (i.e. a forest), an integer $k$}
\KwResult{Yes if $G$ has an $s$-clique (or equivalently an $s$-club) of size $\ge k$. No otherwise.}
\If{ a connected component of $G$ has diameter $\le s$ and size $\ge k$}
{\textbf{Return} Yes}
\eIf{$s\text{ is even}$}
{
\For{$u\in V_G$}
{
\If{$|\overline{B}(u,\dfrac{s}{2})|\ge k$}
{\textbf{Return} Yes}
}
\textbf{Return} No
}
{
\For{$\{u,v\}\in E_G$}
{
\If{$|\overline{B}(u,\dfrac{s-1}{2})\cup\overline{B}(v,\dfrac{s-1}{2})|\ge k$}
{\textbf{Return} Yes}
}
\textbf{Return} No
}

\end{algorithm}

\hfill


\section{\texorpdfstring{$\gamma$}{}-complete subgraph \texorpdfstring{{\sf W[1]}}{}-hard for the degeneracy}\label{sec:gcs_degeneracy} 

In this section, we establish the {\sf W[1]}-hardness of the \gcs{} problem when parameterized by the degeneracy $d$, for any rational $\gamma\in\ ]0,1[$. As a corollary, we obtain that it is also {\sf W[1]}-hard when parameterized by $k$.

\begin{thm}\label{thm:gcs_W1_d}
    {\gcs} is {\sf W[1]}-hard parameterized by $d$ (the degeneracy of the input graph).
\end{thm}

\begin{proof}

We take two integers $a,b>0$ such that $\gamma =\dfrac{a}{b}$.

Let $(G,k)$ be an instance of \clique{}. We can assume that $k$ is of the form $2r(b-a)+2$ with $r>0$ an integer (indeed, $b-a>0$ since $\gamma<1$).

\hfill

\underline{Technicalities:}

\hfill

Recall that we have assumed that $k$ is of the form $2r(b-a)+2$ with $r>0$.

Now, let $R:=r(k+3)>0$ and $p:= Ra - k + 1>0$ (recall that since $\gamma\neq 0$, it holds that $a\neq 0$).

We choose these values because, in our reduction, we will ask for a $\gamma$-complete subgraph of size $k':= 2k - 3 + \binom{k}{2} +p$, and the minimal degree in such a subgraph will necessarily be exactly $d_{\gamma} := k-1+p$. Thus, we want to ensure that such a subgraph is (tightly) $\gamma$-complete, i.e. that $\gamma = \dfrac{d_{\gamma}}{k'-1}$.

Let us check that indeed, $\gamma = \dfrac{d_{\gamma}}{k'-1}$. We will prove first that $d_{\gamma} = Ra$ and then that $k'-1= Rb$, which will imply that $\dfrac{d_{\gamma}}{k'-1} = \dfrac{a}{b}=\gamma$.

\hfill

First, clearly, $d_{\gamma} = k-1+p = k-1 + Ra -k+1 = Ra$ by definitions of $d_{\gamma}$ and $p$.

Second:

\begin{align*}
k'-1 &= 2k-4+\binom{k}{2}+p \\
&= 2k-4+\binom{k}{2}+ Ra-k+1 \\
&= k-3+\dfrac{k(k-1)}{2} +Ra \\
&= \dfrac{k(k+1)}{2}-3 +Ra \\
&= \dfrac{k}{2}(k+1) -3 +Ra \\
&= (r(b-a)+1)(2r(b-a)+3) -3 +Ra \\
&= 2r^2(b-a)^2 + 2r(b-a) + 3r(b-a) +3 -3 +Ra\\
&= r(b-a)(2r(b-a)+2+3) +Ra \\
&= r(b-a)(k+3) +Ra \\
&= R(b-a) + Ra \\
&= Rb
\end{align*}
\hfill

This proves that $\gamma =\dfrac{Ra}{Rb}= \dfrac{d_{\gamma}}{k'-1}$ (note that both $R$ and $b$ are strictly positive).

\hfill

\underline{Construction of $G'$:}

\hfill

First, in order to decrease the degeneracy of the input graph we subdivide each edge by adding a red vertex as done in Figure~\ref{fig:2_degen}.

\begin{figure}[!ht]
    \centering
    \begin{tikzpicture}[scale=0.5]
    \draw[thick,black] (-0.5,-1.5)--(-0.5,-0.5);
    \draw[thick,black] (-0.5,-1.5)--(1.0,-1.0);
    \draw[thick,black] (-0.5,-1.5)--(-1.5,0.5);
    \draw[thick,black] (-0.5,-0.5)--(-0.5,0.5);
    \draw[thick,black] (-0.5,-0.5)--(1.0,-1.0);
    \draw[thick,black] (-0.5,-0.5)--(0.5,0.0);
    \draw[thick,black] (-0.5,-0.5)--(-1.5,0.5);
    \draw[thick,black] (-0.5,0.5)--(0.5,0.0);
    \draw[thick,black] (-0.5,0.5)--(-1.5,0.5);
    \draw[thick,black] (1.0,-1.0)--(0.5,0.0);
    \draw[thick,black] (1.0,-1.0)--(1.0,1.5);
    \draw[thick,black] (0.5,0.0)--(1.0,1.5);
    \draw[thick,black] (-1.5,0.5)--(1.0,1.5);
    
    \filldraw[blue] (-0.5,-1.5) circle (3 pt);
    \filldraw[blue] (-0.5,-0.5) circle (3 pt);
    \filldraw[blue] (-0.5,0.5) circle (3 pt);
    \filldraw[blue] (1.0,-1.0) circle (3 pt);
    \filldraw[blue] (0.5,0.0) circle (3 pt);
    \filldraw[blue] (-1.5,0.5) circle (3 pt);
    \filldraw[blue] (1.0,1.5) circle (3 pt);
    
    \draw[line width=0.5mm, -to] (2.5,0) -- (6,0);
    
    \node at (9.5,0) {
    \begin{tikzpicture}[scale=0.5]
    \draw[thick,black] (-1.0,-3.0)--(-1.0,-2.0);
    \draw[thick,black] (-1.0,-3.0)--(0.6,-2.7);
    \draw[thick,black] (-1.0,-3.0)--(-2.2,-1.1);
    \draw[thick,black] (-1.0,-1.0)--(-1.0,-2.0);
    \draw[thick,black] (-1.0,-1.0)--(-1.0,0.0);
    \draw[thick,black] (-1.0,-1.0)--(0.5,-1.5);
    \draw[thick,black] (-1.0,-1.0)--(0.0,-0.5);
    \draw[thick,black] (-1.0,-1.0)--(-2.0,0.0);
    \draw[thick,black] (-1.0,1.0)--(-1.0,0.0);
    \draw[thick,black] (-1.0,1.0)--(0.0,0.5);
    \draw[thick,black] (-1.0,1.0)--(-2.0,1.0);
    \draw[thick,black] (2.0,-2.0)--(0.6,-2.7);
    \draw[thick,black] (2.0,-2.0)--(0.5,-1.5);
    \draw[thick,black] (2.0,-2.0)--(1.5,-1.0);
    \draw[thick,black] (2.0,-2.0)--(2.2,0.5);
    \draw[thick,black] (1.0,0.0)--(0.0,-0.5);
    \draw[thick,black] (1.0,0.0)--(0.0,0.5);
    \draw[thick,black] (1.0,0.0)--(1.5,-1.0);
    \draw[thick,black] (1.0,0.0)--(1.5,1.5);
    \draw[thick,black] (-3.0,1.0)--(-2.2,-1.1);
    \draw[thick,black] (-3.0,1.0)--(-2.0,0.0);
    \draw[thick,black] (-3.0,1.0)--(-2.0,1.0);
    \draw[thick,black] (-3.0,1.0)--(-0.6,2.2);
    \draw[thick,black] (2.0,3.0)--(2.2,0.5);
    \draw[thick,black] (2.0,3.0)--(1.5,1.5);
    \draw[thick,black] (2.0,3.0)--(-0.6,2.2);
    
    \filldraw[blue] (-1.0,-3.0) circle (3 pt);
    \filldraw[blue] (-1.0,-1.0) circle (3 pt);
    \filldraw[blue] (-1.0,1.0) circle (3 pt);
    \filldraw[blue] (2.0,-2.0) circle (3 pt);
    \filldraw[blue] (1.0,0.0) circle (3 pt);
    \filldraw[blue] (-3.0,1.0) circle (3 pt);
    \filldraw[blue] (2.0,3.0) circle (3 pt);
    \filldraw[red] (-1.0,-2.0) circle (3 pt);
    \filldraw[red] (0.6,-2.7) circle (3 pt);
    \filldraw[red] (-2.2,-1.1) circle (3 pt);
    \filldraw[red] (-1.0,0.0) circle (3 pt);
    \filldraw[red] (0.5,-1.5) circle (3 pt);
    \filldraw[red] (0.0,-0.5) circle (3 pt);
    \filldraw[red] (-2.0,0.0) circle (3 pt);
    \filldraw[red] (0.0,0.5) circle (3 pt);
    \filldraw[red] (-2.0,1.0) circle (3 pt);
    \filldraw[red] (1.5,-1.0) circle (3 pt);
    \filldraw[red] (2.2,0.5) circle (3 pt);
    \filldraw[red] (1.5,1.5) circle (3 pt);
    \filldraw[red] (-0.6,2.2) circle (3 pt);
    \end{tikzpicture}};
    \end{tikzpicture}
    \caption{2-degenerancy transformation of a graph.}
    \label{fig:2_degen}
\end{figure}

\hfill

The rest of the construction of $G'$ is done as follows. $V_{G'}$ contains four types of vertices:

\begin{itemize}

\item The original vertices of $V_G$, which we will call the blue vertices.

\item A red vertex $e_{u,v}$ for each original edge $\{u,v\}\in E_G$ added during the subdivision of the edges. We denote $V_R$ the set formed by the red vertices.

\item $k-3$ new vertices, which we will call the purple vertices. We denote $V_P$ the set formed by the purple vertices.

\item $p$ new vertices, which we will call the yellow vertices. We denote $V_Y$ the set formed by the yellow vertices.

\end{itemize}

We will refer to every vertex that is either yellow or purple as a {\it special} vertex. 

The edges of $G'$ are the following:

\begin{itemize}

\item For every $\{u,v\} \in E_{G}$, the edges of the subdivision, i.e. $\{e_{u,v},u\}$ and $\{e_{u,v},v\}$, exist in $G'$.

\item Every yellow vertex is a universal vertex of $G'$.

\item Every purple vertex is a neighbor of every red vertex.

\item The set of special vertices (yellow and purple) forms a clique (of size $p+k-3$).

\end{itemize}

The graph obtained by this construction is illustrated in Figure~\ref{fig:const_gcs_degen}.

\begin{figure}[!ht]
    \centering
    \begin{tikzpicture}[scale=0.5]
    \draw[thick,blue,dotted] (-0.5,-1.5)--(-0.5,-0.5);
    \draw[thick,blue,dotted] (-0.5,-1.5)--(1.0,-1.0);
    \draw[thick,blue,dotted] (-0.5,-1.5)--(-1.5,0.5);
    \draw[thick,blue,dotted] (-0.5,-0.5)--(-0.5,0.5);
    \draw[thick,blue,dotted] (-0.5,-0.5)--(1.0,-1.0);
    \draw[thick,blue,dotted] (-0.5,-0.5)--(0.5,0.0);
    \draw[thick,blue,dotted] (-0.5,-0.5)--(-1.5,0.5);
    \draw[thick,blue,dotted] (-0.5,0.5)--(0.5,0.0);
    \draw[thick,blue,dotted] (-0.5,0.5)--(-1.5,0.5);
    \draw[thick,blue,dotted] (1.0,-1.0)--(0.5,0.0);
    \draw[thick,blue,dotted] (1.0,-1.0)--(1.0,1.5);
    \draw[thick,blue,dotted] (0.5,0.0)--(1.0,1.5);
    \draw[thick,blue,dotted] (-1.5,0.5)--(1.0,1.5);
    \draw[thick,orange] (12.4,0.0)--(11.9,1.4);
    \draw[thick,orange] (12.4,0.0)--(10.7,2.3);
    \draw[thick,orange] (12.4,0.0)--(9.3,2.3);
    \draw[thick,orange] (12.4,0.0)--(8.1,1.4);
    \draw[thick,orange] (12.4,0.0)--(7.6,0.0);
    \draw[thick,orange] (12.4,0.0)--(8.1,-1.4);
    \draw[thick,orange] (12.4,0.0)--(9.3,-2.3);
    \draw[thick,orange] (12.4,0.0)--(10.7,-2.3);
    \draw[thick,orange] (12.4,0.0)--(11.9,-1.4);
    \draw[thick,orange] (11.9,1.4)--(10.7,2.3);
    \draw[thick,orange] (11.9,1.4)--(9.3,2.3);
    \draw[thick,orange] (11.9,1.4)--(8.1,1.4);
    \draw[thick,orange] (11.9,1.4)--(7.6,0.0);
    \draw[thick,orange] (11.9,1.4)--(8.1,-1.4);
    \draw[thick,orange] (11.9,1.4)--(9.3,-2.3);
    \draw[thick,orange] (11.9,1.4)--(10.7,-2.3);
    \draw[thick,orange] (11.9,1.4)--(11.9,-1.4);
    \draw[thick,orange] (10.7,2.3)--(9.3,2.3);
    \draw[thick,orange] (10.7,2.3)--(8.1,1.4);
    \draw[thick,orange] (10.7,2.3)--(7.6,0.0);
    \draw[thick,orange] (10.7,2.3)--(8.1,-1.4);
    \draw[thick,orange] (10.7,2.3)--(9.3,-2.3);
    \draw[thick,orange] (10.7,2.3)--(10.7,-2.3);
    \draw[thick,orange] (10.7,2.3)--(11.9,-1.4);
    \draw[thick,orange] (9.3,2.3)--(8.1,1.4);
    \draw[thick,orange] (9.3,2.3)--(7.6,0.0);
    \draw[thick,orange] (9.3,2.3)--(8.1,-1.4);
    \draw[thick,orange] (9.3,2.3)--(9.3,-2.3);
    \draw[thick,orange] (9.3,2.3)--(10.7,-2.3);
    \draw[thick,orange] (9.3,2.3)--(11.9,-1.4);
    \draw[thick,orange] (8.1,1.4)--(7.6,0.0);
    \draw[thick,orange] (8.1,1.4)--(8.1,-1.4);
    \draw[thick,orange] (8.1,1.4)--(9.3,-2.3);
    \draw[thick,orange] (8.1,1.4)--(10.7,-2.3);
    \draw[thick,orange] (8.1,1.4)--(11.9,-1.4);
    \draw[thick,orange] (7.6,0.0)--(8.1,-1.4);
    \draw[thick,orange] (7.6,0.0)--(9.3,-2.3);
    \draw[thick,orange] (7.6,0.0)--(10.7,-2.3);
    \draw[thick,orange] (7.6,0.0)--(11.9,-1.4);
    \draw[thick,orange] (8.1,-1.4)--(9.3,-2.3);
    \draw[thick,orange] (8.1,-1.4)--(10.7,-2.3);
    \draw[thick,orange] (8.1,-1.4)--(11.9,-1.4);
    \draw[thick,orange] (9.3,-2.3)--(10.7,-2.3);
    \draw[thick,orange] (9.3,-2.3)--(11.9,-1.4);
    \draw[thick,orange] (10.7,-2.3)--(11.9,-1.4);
    \draw[thick,purple] (12.0,-11.0)--(11.4,-9.6);
    \draw[thick,purple] (12.0,-11.0)--(10.0,-9.0);
    \draw[thick,purple] (12.0,-11.0)--(8.6,-9.6);
    \draw[thick,purple] (12.0,-11.0)--(8.0,-11.0);
    \draw[thick,purple] (12.0,-11.0)--(8.6,-12.4);
    \draw[thick,purple] (12.0,-11.0)--(10.0,-13.0);
    \draw[thick,purple] (12.0,-11.0)--(11.4,-12.4);
    \draw[thick,purple] (11.4,-9.6)--(10.0,-9.0);
    \draw[thick,purple] (11.4,-9.6)--(8.6,-9.6);
    \draw[thick,purple] (11.4,-9.6)--(8.0,-11.0);
    \draw[thick,purple] (11.4,-9.6)--(8.6,-12.4);
    \draw[thick,purple] (11.4,-9.6)--(10.0,-13.0);
    \draw[thick,purple] (11.4,-9.6)--(11.4,-12.4);
    \draw[thick,purple] (10.0,-9.0)--(8.6,-9.6);
    \draw[thick,purple] (10.0,-9.0)--(8.0,-11.0);
    \draw[thick,purple] (10.0,-9.0)--(8.6,-12.4);
    \draw[thick,purple] (10.0,-9.0)--(10.0,-13.0);
    \draw[thick,purple] (10.0,-9.0)--(11.4,-12.4);
    \draw[thick,purple] (8.6,-9.6)--(8.0,-11.0);
    \draw[thick,purple] (8.6,-9.6)--(8.6,-12.4);
    \draw[thick,purple] (8.6,-9.6)--(10.0,-13.0);
    \draw[thick,purple] (8.6,-9.6)--(11.4,-12.4);
    \draw[thick,purple] (8.0,-11.0)--(8.6,-12.4);
    \draw[thick,purple] (8.0,-11.0)--(10.0,-13.0);
    \draw[thick,purple] (8.0,-11.0)--(11.4,-12.4);
    \draw[thick,purple] (8.6,-12.4)--(10.0,-13.0);
    \draw[thick,purple] (8.6,-12.4)--(11.4,-12.4);
    \draw[thick,purple] (10.0,-13.0)--(11.4,-12.4);
    
    \draw[thick,black] (1.7,-1.2)--(7.4,1.5);
    \draw[thick,black] (1.7,-1.2)--(7.0,0.0);
    \draw[thick,black] (1.7,-1.2)--(7.4,-1.5);
    \draw[thick,black] (2.0,0.0)--(7.4,1.5);
    \draw[thick,black] (2.0,0.0)--(7.0,0.0);
    \draw[thick,black] (2.0,0.0)--(7.4,-1.5);
    \draw[thick,black] (1.7,1.2)--(7.4,1.5);
    \draw[thick,black] (1.7,1.2)--(7.0,0.0);
    \draw[thick,black] (1.7,1.2)--(7.4,-1.5);
    
    \draw[thick,black] (2.4,-10.2)--(7.1,-0.9);
    \draw[thick,black] (2.4,-10.2)--(8.0,-2.2);
    \draw[thick,black] (2.4,-10.2)--(9.4,-2.9);
    \draw[thick,black] (1.7,-9.2)--(7.1,-0.9);
    \draw[thick,black] (1.7,-9.2)--(8.0,-2.2);
    \draw[thick,black] (1.7,-9.2)--(9.4,-2.9);
    \draw[thick,black] (0.5,-8.6)--(7.1,-0.9);
    \draw[thick,black] (0.5,-8.6)--(8.0,-2.2);
    \draw[thick,black] (0.5,-8.6)--(9.4,-2.9);
    
    \draw[thick,black] (11.2,-8.8)--(8.5,-2.6);
    \draw[thick,black] (11.2,-8.8)--(10.0,-3.0);
    \draw[thick,black] (11.2,-8.8)--(11.5,-2.6);
    \draw[thick,black] (10.0,-8.5)--(8.5,-2.6);
    \draw[thick,black] (10.0,-8.5)--(10.0,-3.0);
    \draw[thick,black] (10.0,-8.5)--(11.5,-2.6);
    \draw[thick,black] (8.8,-8.8)--(8.5,-2.6);
    \draw[thick,black] (8.8,-8.8)--(10.0,-3.0);
    \draw[thick,black] (8.8,-8.8)--(11.5,-2.6);
    
    \draw[thick,black] (7.8,-9.8)--(2.2,-12.2);
    \draw[thick,black] (7.8,-9.8)--(2.5,-11.0);
    \draw[thick,black] (7.8,-9.8)--(2.2,-9.8);
    \draw[thick,black] (7.5,-11.0)--(2.2,-12.2);
    \draw[thick,black] (7.5,-11.0)--(2.5,-11.0);
    \draw[thick,black] (7.5,-11.0)--(2.2,-9.8);
    \draw[thick,black] (7.8,-12.2)--(2.2,-12.2);
    \draw[thick,black] (7.8,-12.2)--(2.5,-11.0);
    \draw[thick,black] (7.8,-12.2)--(2.2,-9.8);

    \filldraw[blue] (-0.5,-1.5) circle (3 pt);
    \filldraw[blue] (-0.5,-0.5) circle (3 pt);
    \filldraw[blue] (-0.5,0.5) circle (3 pt);
    \filldraw[blue] (1.0,-1.0) circle (3 pt);
    \filldraw[blue] (0.5,0.0) circle (3 pt);
    \filldraw[blue] (-1.5,0.5) circle (3 pt);
    \filldraw[blue] (1.0,1.5) circle (3 pt);
    \filldraw[red] (1.8,-11.0) circle (3 pt);
    \filldraw[red] (1.5,-10.2) circle (3 pt);
    \filldraw[red] (1.0,-9.6) circle (3 pt);
    \filldraw[red] (0.2,-9.3) circle (3 pt);
    \filldraw[red] (-0.6,-9.4) circle (3 pt);
    \filldraw[red] (-1.3,-9.8) circle (3 pt);
    \filldraw[red] (-1.7,-10.6) circle (3 pt);
    \filldraw[red] (-1.7,-11.4) circle (3 pt);
    \filldraw[red] (-1.3,-12.2) circle (3 pt);
    \filldraw[red] (-0.6,-12.6) circle (3 pt);
    \filldraw[red] (0.2,-12.7) circle (3 pt);
    \filldraw[red] (1.0,-12.4) circle (3 pt);
    \filldraw[red] (1.5,-11.8) circle (3 pt);
    \filldraw[orange] (12.4,0.0) circle (3 pt);
    \filldraw[orange] (11.9,1.4) circle (3 pt);
    \filldraw[orange] (10.7,2.3) circle (3 pt);
    \filldraw[orange] (9.3,2.3) circle (3 pt);
    \filldraw[orange] (8.1,1.4) circle (3 pt);
    \filldraw[orange] (7.6,0.0) circle (3 pt);
    \filldraw[orange] (8.1,-1.4) circle (3 pt);
    \filldraw[orange] (9.3,-2.3) circle (3 pt);
    \filldraw[orange] (10.7,-2.3) circle (3 pt);
    \filldraw[orange] (11.9,-1.4) circle (3 pt);
    \filldraw[purple] (12.0,-11.0) circle (3 pt);
    \filldraw[purple] (11.4,-9.6) circle (3 pt);
    \filldraw[purple] (10.0,-9.0) circle (3 pt);
    \filldraw[purple] (8.6,-9.6) circle (3 pt);
    \filldraw[purple] (8.0,-11.0) circle (3 pt);
    \filldraw[purple] (8.6,-12.4) circle (3 pt);
    \filldraw[purple] (10.0,-13.0) circle (3 pt);
    \filldraw[purple] (11.4,-12.4) circle (3 pt);
    
    \draw[-{Straight Barb[right]},line width=0.5mm,black] (0.2,-8.5)--(0.2,-2.5);
    \node[rotate=90] at (-0.8,-5.5) {$\deg_{G}(v)$};
    \draw[-{Straight Barb[right]},line width=0.5mm,black] (-0.2,-2.5)--(-0.2,-8.5);
    \node[rotate=-90] at (0.8,-5.5) {$2$};

    \draw[thick,blue] (0.0,0.0) ellipse (2.0cm and 2.5cm);
    \node[blue] at (-3.1,2.8) {$V_{G},$};
    \node[blue] at (-3.1,2.1) {$|V_{G}|=n$};
    
    \draw[thick,red] (0.0,-11.0) ellipse (2.5cm and 2.5cm);
    \node[red] at (-3,-12.8) {$V_R,$};
    \node[red] at (-3,-13.5) {$|V_R|=m$};
    
    \draw[thick,orange] (10.0,0.0) ellipse (3.0cm and 3.0cm);
    \node[orange] at (14.5,3.5) {$V_Y,$};
    \node[orange] at (14.5,2.8) {$|V_Y|=p$};
    
    \draw[thick,purple] (10.0,-11.0) ellipse (2.5cm and 2.5cm);
    \node[purple] at (13.2,-13.2) {$V_P,$};
    \node[purple] at (13.2,-13.9) {$|V_P|=k-3$};
    \end{tikzpicture}
    \caption{Construction of $G'$. 9 edges between sets of vertices represent a complete linkage. The double arrow represent that each blue vertex $v$ has $\deg_{G}(v)$ red neighbors and each red vertex has 2 blue neighbors.}
    \label{fig:const_gcs_degen}
\end{figure}
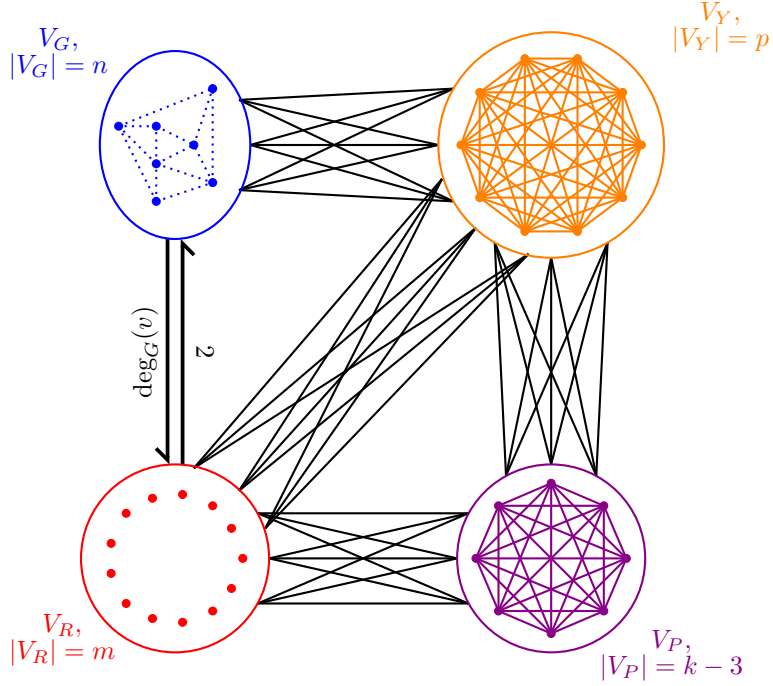

\hfill

Note that $G'$ can be constructed in time $n^{O(1)}$, where $n=|V_{G}|$.

\hfill

\underline{Soundness of the reduction:}

\hfill

Recall that we defined $k':= 2k-3+ \binom{k}{2}+ p = k+ \binom{k}{2} + |V_Y| + |V_P| $. We prove that $(G,k)$ is a yes-instance of {\clique} if, and only if, $(G',k')$ is a yes-instance of \gcs .

But first, the following claim proves useful properties on any $\gamma$-complete-subgraph of $G'$ of size $\ge k'$.

\begin{claim}\label{claim:exact card}

Let $S$ be a $\gamma$-complete subgraph of size $\ge k'$ of $G'$. Then:

\begin{enumerate}

\item $S$ contains a red vertex.

\item $|S|=k'$.

\item If a red vertex $e_{u,v}$ is in $S$, then the two blue vertices $u$ and $v$ are also in $S$.

\item $S$ contains every special vertices

\item If a blue vertex $u$ is in $S$, then it has at least $k-1$ red neighbors in $S$.

\end{enumerate}

\end{claim}

\begin{proofclaim}

\begin{enumerate}

\item Assume by contradiction that $S$ contains no red vertex. Since there are exactly $p$ special vertices, and that $p < k' \le |S|$, $S$ contains at least one non-special vertex $u$, i.e.: $u$ is either blue or red: $u$ is therefore blue. On the one hand, since $S$ is a $\gamma$-complete graph, $\deg_S(u) \ge \gamma(|S'|-1) \ge \gamma(k'-1) = d_{\gamma}$. On the other hand, the neighbors of $u$ in $S$ are neither blue, nor purple (because a blue vertex is never adjacent to a blue/purple vertex) nor red (because we assumed that $S$ does not contain red vertices): so $\deg_S(u)\le p$ (the only possible neighbors of $u$ in $S$ are the $p$ yellow vertices), which is absurd because $p<d_{\gamma}$ (recall that $d_{\gamma}=k-1+p$).

\item Now that we have proven that $S$ contains thus a red vertex $e_{u,v}$, assume by contradiction that $|S|>k'$. On the one hand, since $S$ is a $\gamma$-complete subgraph, we have $\deg_S(e_{u,v})\ge \gamma (|S|-1)  > \gamma (k'-1)=d_{\gamma}$. On the other hand, $\deg_{G'}(e_{u,v}) = 2 + (k-3) + p = d_{\gamma}$ (a red vertex has exactly $2$ blue neighbors, $k-3$ purple neighbors, and $p$ yellow neighbors). We conclude that $\deg_S(e_{u,v})> \deg_{G'}(e_{u,v})$, which is absurd (the degree of a vertex in an induced subgraph can not be strictly greater than its degree in the original graph). This proves that $|S|=k'$.

\item For every red vertex $e_{u,v}$ of $S$, the similar remark shows that we have $\displaystyle{\deg_S(e_{u,v}) \ge \gamma (k'-1) = d_{\gamma} = \deg_{G'}(e_{u,v}) \ge \deg_S(e_{u,v})}$ holds, and therefore, $\displaystyle{\deg_S(e_{u,v})=\deg_{G'}(e_{u,v})}$, i.e. every neighbor of $e_{u,v}$ in $G'$ is in $S$: in particular, $u$ and $v$ are in $S$.

\item Since we know from 1. that $S$ has at least one red vertex $e_{u,v}$, the reasoning of 3. above also shows that every special vertex is in $S$ (because the special vertices are all adjacent to $e_{u,v}$).

\item Every blue vertex in $S$ must have at least $d_{\gamma}=k-1+p$ neighbors in $S$. Since it has already $p$ neighbors within the special vertices, it must have at least $k-1$ other neighbors, which must be red since any two blue vertices are not adjacent in $G'$.

\end{enumerate}

\end{proofclaim}

We are ready to prove that $(G,k)$ is a yes-instance of \clique{} if, and only if, $(G',k')$ is a yes-instance of \gcs :

\begin{itemize}
    \item[$\implies$] Assume that there is a clique $K\subseteq V_G$ of size $k$ in $G$. 

    Take in $S$:

    \begin{itemize}
        
        \item The $k$ blue vertices of $K$.

        \item The $\binom{k}{2}$ red vertices of the form $e_{u,v}$ for $(u,v)\in K^2$ with $u\neq v$ (which all exists since $K$ is a clique in $G$).

        \item All $p+k-3$ special vertices.

    \end{itemize}

    We have indeed $S\subseteq V_{G'}$ and $|S|=k+\binom{k}{2} + p + k - 3 = k'\ge k'$. 

    We verify that $S$ is a $\gamma$-complete subgraph of $G'$:

    \begin{itemize}

        \item Every blue vertex $u\in S$ (i.e. $u\in K$) is adjacent to the $k-1$ red vertices of the form $e_{u,v}$ for $v\in K\setminus \{u\}$. These red vertices are indeed in $S$. In addition, $u$ is also adjacent to each of the $p$ yellow vertices. This proves that $\deg_S(u) \geq  k-1+p = d_{\gamma} = \gamma (k'-1) = \gamma(|S|-1)$.

        \item Every red vertex $e_{u,v}$ in $S$ is adjacent to the $2$ blue vertices $u$ and $v$. These $2$ vertices are indeed in $S$. In addition, $e_{u,v}$ is adjacent to the $k-3$ purple vertices, and to the $d_{\gamma}-k+1$ yellow vertices. Thus proves that $\deg_S(e_{u,v})\ge 2+ (k-3) + p = d_{\gamma} = \gamma (|S|-1)$.

        \item Every special vertex is adjacent to every other special and vertex, and to every red vertex in $S$: thus, as $\binom{k}{2}\geq 3$, their degree is at least $\displaystyle{p + (k-3) -1 + \binom{k}{2} \ge d_{\gamma}}$.

    \end{itemize}

    We have proven that $S$ is a $\gamma$-complete subgraph of size $\ge k'$ of $G'$.
    
    \item[$\impliedby$] Assume that there is a $\gamma$-complete subgraph $S$ of size $\geq k'$ of $G'$.

    Let $V_K$ be the set of blue vertices in $S$ (i.e. $V_K=S\cap V_G$), and $E_K$ be the set of edges $\{u,v\}$ of $G$ that are such that $e_{u,v}\in S$ (formally, we have $\displaystyle{E_K=\{ \{u,v\}\in E_G \mid e_{u,v}\in S\}}$). By Claim~\ref{claim:exact card}.3, for any red vertex $e_{u,v}\in S$, it holds that $\{u,v\}\in (V_K)^2$. In other words, we have that $K=(V_K,E_K)$ is a subgraph (not necessarily induced) of $G$. We will prove that $K$ is an induced clique of size $k$ in $G$.

    By Claim~\ref{claim:exact card}.2 and \ref{claim:exact card}.4, we have $|S|=k'=2k-3+\binom{k}{2}+p$, and that $S$ contains all $p+k-3$ special vertices. It follows that $|V_K|+|E_K|=k+\binom{k}{2}$. We will prove that $|V_K|=k$ and that $|E_K|=\binom{k}{2}$.

    Assume by contradiction that $|V_K|<k$. Then the graph $K$ which has $|V_K|$ vertices has at most $\binom{|V_K|}{2} < \binom{k}{2}$ edges, i.e. $|E_K|<\binom{k}{2}$. We have a contradiction with $|V_K|+|E_K|=k+\binom{k}{2}$.

    Assume by contradiction that $|V_K|>k$. Then, by Claim~\ref{claim:exact card}.5 the degree of any vertex in the graph $K$ is at least $k-1$. Thus, we have by Lemma~\ref{lem:sum_degree} that $2|E_K| = \sum\limits_{u\in V_K} \deg_K(u) \ge \sum\limits_{u\in V_K} (k-1) = |V_K|(k-1)> k(k-1)$. It follows that $|E_K|> \dfrac{k(k-1)}{2}=\binom{k}{2}$. We have again a contradiction with $|V_K|+|E_K|=k+\binom{k}{2}$.

    This proves that $|V_K|=k$, and thus that $|E_K|=\binom{k}{2}$. Therefore, $K$ is a subgraph of $G$ with $k$ vertices and $\binom{k}{2}$ edges: it is an induced clique of size $k$ in $G$.
    
\end{itemize}

\underline{Degeneracy of $G'$:}

\hfill

Finally, we prove that the degeneracy of $G'$ is at most $k-1+p$, and thus, depends only on $k$, by giving an elimination order:

\begin{itemize}

    \item First, remove the red vertices: they have degree $2+k-3+p=k-1+p$.

    \item Second, remove the blue vertices: they are now only adjacent to the $p$ yellow vertices.

    \item Finally, remove the $p+k-3$ special vertices.

\end{itemize}

Since \clique{} is {\sf W[1]}-hard parameterized by $k$, this completes the proof that \gcs{} is {\sf W[1]}-hard parameterized by $d$, the degeneracy of the input graph.

\end{proof}

As an immediate corollary of Theorem~\ref{thm:gcs_W1_d}, by comparing the parameters $k$ and $d$ on graphs that have a $\gamma$-complete-subgraphs, we obtained the {\sf W[1]}-hardness of \gcs{} when parameterized by $k$.

\begin{cor}\label{cor:gcs_W1_k}

{\gcs} is {\sf W[1]}-hard parameterized by $k$.

\end{cor}

\begin{proof}

This is due to Remark~\ref{rem:W1-hard_comparaison} and the fact that {\gcs} is trivial (there are only no-instances) as soon as $d< \gamma(k-1)$ (where $d$ is the degeneracy of the input graph). Indeed, if $S$ is a subset of at least $k$ vertices of a graph $G$ that forms a $\gamma$-complete-subgraph of $G$, it has a vertex of degree at most $d$ by definition of $d$. Using that $S$ is a $\gamma$-complete-subgraph, we get $d\ge \gamma(|S|-1) \ge \gamma(k-1)$. Alternatively, one could notice that in the reduction in the proof Theorem~\ref{thm:gcs_W1_d}, $k'$ depends only on $k$, and is thus also a reduction to \gcs{} parameterized by $k$.

\end{proof}

Notice also that unless {\sf P=NP} the result of {\sf W[1]}-hardness of \gcs{}, presented in Theorem~\ref{thm:gcs_W1_d}, cannot be improved in a result of para-{\sf NP}-hardness. Indeed, \gcs{} is clearly {\sf XP} when parameterized by $d$. For $(G,k)$ an instance of \gcs{} and $d$ the degeneracy of $G$, if $d<\gamma(k-1)$ then $(G,k)$ is a no-instance. Otherwise, one just have to test, for every $S \subseteq V_{G}$ such that $k \leq |S| \leq \frac{1}{\gamma}d+1$, if $S$ is a $\gamma$-complete-subgraph of $G$. This process can be performed in {\sf XP} time, more precisely in time $O(n^3 \times n^{\frac{1}{\gamma}d})$.


\section{\texorpdfstring{$\gamma$}{}-complete subgraph \texorpdfstring{{\sf W[1]}}{}-hard for \texorpdfstring{$\ell = n-k$}{}}\label{sec:gcs_l} 

In this section, we establish that \gcs{} is also {\sf W[1]}-hard when parameterized by $\ell:=n-k$, the number of vertices outside the $\gamma$-complete-subgraph.

This result comes as a surprise, since not only is \clique{} {\sf FPT} parameterized by $\ell$ \cite{komusiewicz2016multivariate}, but so is the {\sc $s$-Plex} problem \cite{komusiewicz2009isolation}: that asks, given a graph $G$ and an integer $k$, if there exists $S\subseteq V_G$ with $|S|\ge k$ and $\forall u\in S, \deg_S(u)\ge |S|-s$. The only difference between {\sc $s$-Plex} and \gcs{} is that the constraint on the number of non-neighbors depends on the size of $S$ in the latter.

\begin{thm}\label{thm:gCS_W1_l}
{\gcs} is {\sf W[1]}-hard parameterized by $\ell=n-k$ where $k$ is the size of the $\gamma$-complete subgraph.
\end{thm}

\begin{proof}

Let $(G,k)$ be an instance of \clique{}. We assume without loss of generality that $k > \dfrac{2}{\gamma}$. We construct $(G',k')$ an equivalent instance of \gcs{} as follows.

\hfill

\underline{Technicalities:}

\hfill

For the sake of our reduction, we will need an integer $N$ that satisfies some useful properties.

\begin{claim}\label{claim:tech_N}
There exists $N$ an integer such that:
\begin{enumerate}
    \item $\displaystyle N > \dfrac{1}{1-\gamma}(\gamma n+\gamma \overline{m}+1)$,
    \item $\displaystyle \left\lceil\gamma(N+n+\overline{m})\right\rceil < \left\lceil\gamma(N+n+\overline{m}+1)\right\rceil$,
    \item $N > (k+1)(k+\overline{m})$.
    \item $N$ is polynomial in $n$.
\end{enumerate}
\end{claim}

\begin{proof} (Claim~\ref{claim:tech_N})

Note that 2 is true at least once every $\lceil\dfrac{1}{\gamma}\rceil$ integers. Recall that $k$ is bounded by $n$, hence, an $N$ verifying points 1 to 3 always exists. Moreover, the smallest possible $N$ verifies:
$$ N < \max \left(\dfrac{1}{1-\gamma}(\gamma n+\gamma \overline{m}+1), 
\quad (k+1)(k+\overline{m}) \right) + \left\lceil\dfrac{1}{\gamma}\right\rceil +1. $$
\end{proof}

Let $\displaystyle d_{\gamma} := \left\lceil\gamma(N+n+\overline{m}) \right\rceil$. Note that $N \geq d_{\gamma}$: it is indeed a direct consequence of Claim~\ref{claim:tech_N}.1.

\hfill

\underline{Construction of $G'$:}

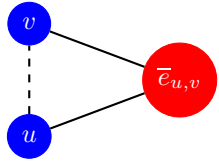
\begin{figure}[!ht]
    \centering
    \begin{tikzpicture}[scale=0.5]
    \node[blue,draw,circle,fill=blue] (u) at (0,0) {\textcolor{white}{$u$}};
    \node[blue,draw,circle,fill=blue] (v) at (0,3) {\textcolor{white}{$v$}};
    \node[red,draw,circle,fill=red] (euv) at (4,1.5) {\textcolor{white}{$\overline{e}_{u,v}$}};
    \draw[thick,dashed] (u) -- (v);
    \draw[thick] (u) -- (euv);
    \draw[thick] (v) -- (euv);
    \end{tikzpicture}
    \caption{Widget performed on the non-edges of $G$.}
    \label{fig:widget_gcs_ell}
\end{figure}

\hfill

\noindent We construct $V_{G'}$ with the following vertices.

\begin{itemize}
    \item The original vertices of $V_{G}$ which we call the blue vertices.

    \item For each $\{u,v\} \in \overline{E_{G}}$ the vertex $\overline{e}_{u,v}$ which we call a red vertex. We denote $V_R$ the set formed by the red vertices.

    \item $k+1$ new purple vertices forming the set $V_P$.

    \item $N$ new yellow vertices forming the set $V_Y$.
\end{itemize}

\noindent The vertices of $V_{G'}$ are connected by the following edges:

\begin{itemize}
    \item The original edges of $E_{G}$.

    \item For each $\{u,v\} \in \overline{E_{G}}$, the edges $\{u,\overline{e}_{u,v}\}$, $\{v,\overline{e}_{u,v}\}$ as done in Figure~\ref{fig:widget_gcs_ell}.

    \item All possible edges between the red and purple vertices, i.e. the bipartite graph induced by $(V_R,V_P)$ is complete.

    \item All possible edges between the purple vertices, i.e. $V_P$ is a clique of $G'$.

    \item All possible edges between yellow vertices, i.e. $V_Y$ is a clique of $G'$.

    \item All possible edges between the blue and yellow vertices, i.e. the bipartite graph induced by $(V_G,V_Y)$ is complete.

    \item For each $\overline{e}_{u,v} \in V_R$: $d_{\gamma}-k-2$ neighbors in the yellow vertices $V_Y$ ($d_{\gamma}-k-2$ is indeed positive).

    \item For each $v \in V_P$: $d_{\gamma}-\overline{m}-k$ neighbors in the yellow vertices $V_Y$ ($d_{\gamma}-\overline{m}-k$ is indeed positive). The first purple with the $d_{\gamma}-\overline{m}-k$ first yellow vertices, the second purple with the next $d_{\gamma}-\overline{m}-k$ yellow vertices and so on. More formally the $i$-th purple vertex is linked to the yellow vertices with indices from $i \times (d_{\gamma}-\overline{m}-k) \mod N$ to $(i+1) \times (d_{\gamma}-\overline{m}-k)-1 \mod N$.
    
    Note that by definition  of $d_{\gamma} := \lceil \gamma(N+n+\overline{m}) \rceil$, we have that $d_{\gamma} \geq \gamma N$. Since $k > \dfrac{2}{\gamma}$ it holds that $(k+1)d_{\gamma} > 2N$, then using Claim~\ref{claim:tech_N}.3 we obtain that $(k+1)(d_{\gamma}-\overline{m}-k) > 2N - (k+1)(\overline{m}+k) > N$. This proves that each yellow vertex has at least a purple neighbor.
\end{itemize}

The graph obtained by this construction is illustrated in Figure~\ref{fig:const_gcs_ell}.

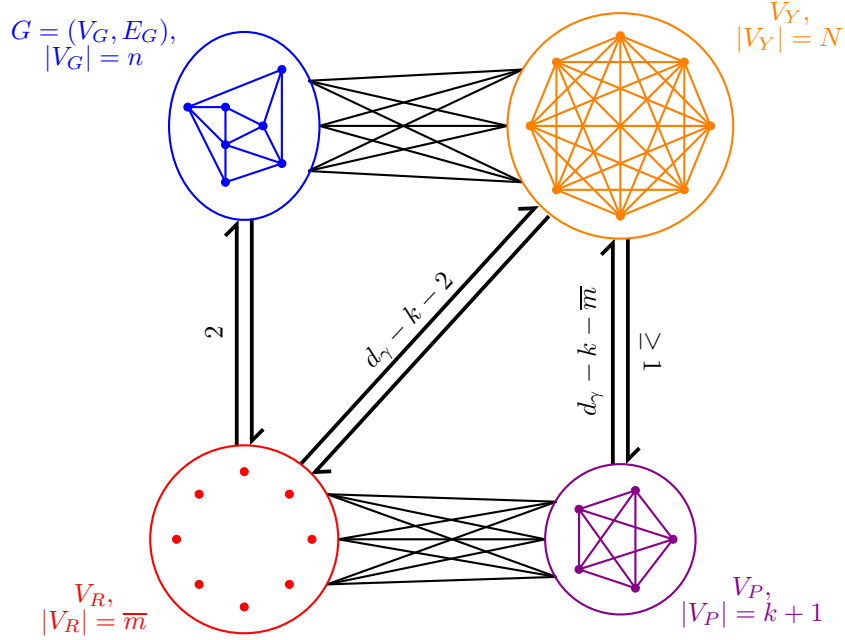
\begin{figure}[!ht]
    \centering
    \begin{tikzpicture}[scale=0.5]
    \draw[thick,blue] (-0.5,-1.5)--(-0.5,-0.5);
    \draw[thick,blue] (-0.5,-1.5)--(1.0,-1.0);
    \draw[thick,blue] (-0.5,-1.5)--(-1.5,0.5);
    \draw[thick,blue] (-0.5,-0.5)--(-0.5,0.5);
    \draw[thick,blue] (-0.5,-0.5)--(1.0,-1.0);
    \draw[thick,blue] (-0.5,-0.5)--(0.5,0.0);
    \draw[thick,blue] (-0.5,-0.5)--(-1.5,0.5);
    \draw[thick,blue] (-0.5,0.5)--(0.5,0.0);
    \draw[thick,blue] (-0.5,0.5)--(-1.5,0.5);
    \draw[thick,blue] (1.0,-1.0)--(0.5,0.0);
    \draw[thick,blue] (1.0,-1.0)--(1.0,1.5);
    \draw[thick,blue] (0.5,0.0)--(1.0,1.5);
    \draw[thick,blue] (-1.5,0.5)--(1.0,1.5);
    \draw[thick,purple] (11.4,-11.0)--(10.4,-9.7);
    \draw[thick,purple] (11.4,-11.0)--(8.9,-10.2);
    \draw[thick,purple] (11.4,-11.0)--(8.9,-11.8);
    \draw[thick,purple] (11.4,-11.0)--(10.4,-12.3);
    \draw[thick,purple] (10.4,-9.7)--(8.9,-10.2);
    \draw[thick,purple] (10.4,-9.7)--(8.9,-11.8);
    \draw[thick,purple] (10.4,-9.7)--(10.4,-12.3);
    \draw[thick,purple] (8.9,-10.2)--(8.9,-11.8);
    \draw[thick,purple] (8.9,-10.2)--(10.4,-12.3);
    \draw[thick,purple] (8.9,-11.8)--(10.4,-12.3);
    
    \draw[thick,orange] (12.4,0.0)--(11.7,1.7);
    \draw[thick,orange] (12.4,0.0)--(10.0,2.4);
    \draw[thick,orange] (12.4,0.0)--(8.3,1.7);
    \draw[thick,orange] (12.4,0.0)--(7.6,0.0);
    \draw[thick,orange] (12.4,0.0)--(8.3,-1.7);
    \draw[thick,orange] (12.4,0.0)--(10.0,-2.4);
    \draw[thick,orange] (12.4,0.0)--(11.7,-1.7);
    \draw[thick,orange] (11.7,1.7)--(10.0,2.4);
    \draw[thick,orange] (11.7,1.7)--(8.3,1.7);
    \draw[thick,orange] (11.7,1.7)--(7.6,0.0);
    \draw[thick,orange] (11.7,1.7)--(8.3,-1.7);
    \draw[thick,orange] (11.7,1.7)--(10.0,-2.4);
    \draw[thick,orange] (11.7,1.7)--(11.7,-1.7);
    \draw[thick,orange] (10.0,2.4)--(8.3,1.7);
    \draw[thick,orange] (10.0,2.4)--(7.6,0.0);
    \draw[thick,orange] (10.0,2.4)--(8.3,-1.7);
    \draw[thick,orange] (10.0,2.4)--(10.0,-2.4);
    \draw[thick,orange] (10.0,2.4)--(11.7,-1.7);
    \draw[thick,orange] (8.3,1.7)--(7.6,0.0);
    \draw[thick,orange] (8.3,1.7)--(8.3,-1.7);
    \draw[thick,orange] (8.3,1.7)--(10.0,-2.4);
    \draw[thick,orange] (8.3,1.7)--(11.7,-1.7);
    \draw[thick,orange] (7.6,0.0)--(8.3,-1.7);
    \draw[thick,orange] (7.6,0.0)--(10.0,-2.4);
    \draw[thick,orange] (7.6,0.0)--(11.7,-1.7);
    \draw[thick,orange] (8.3,-1.7)--(10.0,-2.4);
    \draw[thick,orange] (8.3,-1.7)--(11.7,-1.7);
    \draw[thick,orange] (10.0,-2.4)--(11.7,-1.7);
    
    \draw[thick,black] (2.2,-12.2)--(8.3,-10.0);
    \draw[thick,black] (2.2,-12.2)--(8.0,-11.0);
    \draw[thick,black] (2.2,-12.2)--(8.3,-12.0);
    \draw[thick,black] (2.5,-11.0)--(8.3,-10.0);
    \draw[thick,black] (2.5,-11.0)--(8.0,-11.0);
    \draw[thick,black] (2.5,-11.0)--(8.3,-12.0);
    \draw[thick,black] (2.2,-9.8)--(8.3,-10.0);
    \draw[thick,black] (2.2,-9.8)--(8.0,-11.0);
    \draw[thick,black] (2.2,-9.8)--(8.3,-12.0);
    
    \draw[thick,black] (1.7,-1.2)--(7.4,1.5);
    \draw[thick,black] (1.7,-1.2)--(7.0,0.0);
    \draw[thick,black] (1.7,-1.2)--(7.4,-1.5);
    \draw[thick,black] (2.0,0.0)--(7.4,1.5);
    \draw[thick,black] (2.0,0.0)--(7.0,0.0);
    \draw[thick,black] (2.0,0.0)--(7.4,-1.5);
    \draw[thick,black] (1.7,1.2)--(7.4,1.5);
    \draw[thick,black] (1.7,1.2)--(7.0,0.0);
    \draw[thick,black] (1.7,1.2)--(7.4,-1.5);

    \filldraw[blue] (-0.5,-1.5) circle (3 pt);
    \filldraw[blue] (-0.5,-0.5) circle (3 pt);
    \filldraw[blue] (-0.5,0.5) circle (3 pt);
    \filldraw[blue] (1.0,-1.0) circle (3 pt);
    \filldraw[blue] (0.5,0.0) circle (3 pt);
    \filldraw[blue] (-1.5,0.5) circle (3 pt);
    \filldraw[blue] (1.0,1.5) circle (3 pt);
    \filldraw[red] (1.8,-11.0) circle (3 pt);
    \filldraw[red] (1.2,-9.8) circle (3 pt);
    \filldraw[red] (0.0,-9.2) circle (3 pt);
    \filldraw[red] (-1.2,-9.8) circle (3 pt);
    \filldraw[red] (-1.8,-11.0) circle (3 pt);
    \filldraw[red] (-1.2,-12.2) circle (3 pt);
    \filldraw[red] (-0.0,-12.8) circle (3 pt);
    \filldraw[red] (1.2,-12.2) circle (3 pt);
    \filldraw[purple] (11.4,-11.0) circle (3 pt);
    \filldraw[purple] (10.4,-9.7) circle (3 pt);
    \filldraw[purple] (8.9,-10.2) circle (3 pt);
    \filldraw[purple] (8.9,-11.8) circle (3 pt);
    \filldraw[purple] (10.4,-12.3) circle (3 pt);
    \filldraw[orange] (12.4,0.0) circle (3 pt);
    \filldraw[orange] (11.7,1.7) circle (3 pt);
    \filldraw[orange] (10.0,2.4) circle (3 pt);
    \filldraw[orange] (8.3,1.7) circle (3 pt);
    \filldraw[orange] (7.6,0.0) circle (3 pt);
    \filldraw[orange] (8.3,-1.7) circle (3 pt);
    \filldraw[orange] (10.0,-2.4) circle (3 pt);
    \filldraw[orange] (11.7,-1.7) circle (3 pt);
    
    \draw[-{Straight Barb[left]},line width=0.5mm,black] (0.2,-2.5)--(0.2,-8.5);
    \draw[-{Straight Barb[left]},line width=0.5mm,black] (-0.2,-8.5)--(-0.2,-2.5);
    \node[rotate=89] at (-0.8,-5.5) {$2$};
    
    \draw[-{Straight Barb[left]},line width=0.5mm,black] (1.5,-9.0)--(7.8,-2.1);
    \node[rotate=47] at (4.4,-5.2) {$d_{\gamma}-k-2$};
    \draw[-{Straight Barb[left]},line width=0.5mm,black] (8.1,-2.4)--(1.8,-9.3);
    
    \draw[-{Straight Barb[left]},line width=0.5mm,black] (9.8,-9.0)--(9.8,-3.0);
    \node[rotate=90] at (9.2,-6.0) {$d_{\gamma}-k-\overline{m}$};
    \draw[-{Straight Barb[left]},line width=0.5mm,black] (10.2,-3.0)--(10.2,-9.0);
    \node[rotate=-90] at (10.8,-6.0) {$\geq 1$};

    \draw[thick,blue] (0.0,0.0) ellipse (2.0cm and 2.5cm);
    \node[blue] at (-4,2.5) {$G=(V_{G},E_{G}),$};
    \node[blue] at (-4,1.8) {$|V_{G}|=n$};
    
    \draw[thick,red] (0.0,-11.0) ellipse (2.5cm and 2.5cm);
    \node[red] at (-4,-12.5) {$V_R,$};
    \node[red] at (-4,-13.2) {$|V_R|=\overline{m}$};
    
    \draw[thick,purple] (10.0,-11.0) ellipse (2.0cm and 2.0cm);
    \node[purple] at (13.5,-12.3) {$V_P,$};
    \node[purple] at (13.5,-13) {$|V_P|=k+1$};
    
    \draw[thick,orange] (10.0,0.0) ellipse (3.0cm and 3.0cm);
    \node[orange] at (14.5,3.0) {$V_Y,$};
    \node[orange] at (14.5,2.3) {$|V_Y|=N$};
    
    \end{tikzpicture}
    \caption{Construction of $G'$. 9 edges between sets of vertices represent a complete linkage. A double arrow $A \overset{a}{\underset{b}{\rightleftharpoons}} B$ represents that the vertices in $A$ have $a$ neighbors in $B$ and the vertices in $B$ have $b$ neighbors in $A$, if either $a$ or $b$ is not mentioned it means that this number is irrelevant to the proof.}
    \label{fig:const_gcs_ell}
\end{figure}

Let $k' := N+n+\overline{m}+1$. Notice that $d_{\gamma}=\lceil \gamma (k'-1)\rceil$.

Note that $G'$ can be constructed in time $n^{O(1)}$.

\hfill

\underline{Soundness of the reduction:}

\hfill

First, the following claim proves useful properties on any $\gamma$-complete-subgraph of $G'$ of size $\ge k'$.

\begin{claim}\label{claim:gcs_ell}

Let $S$ be a $\gamma$-complete subgraph of $G'$ of size $\ge k'$. Then:

\begin{enumerate}

    \item $S$ contains every non-blue vertex.

    \item $|S|=k'$.

    \item For every $\{u,v\}\in \overline{E_G}$, either $u$ or $v$ is in $S$.

\end{enumerate}

\end{claim}

\begin{proofclaim}

\begin{enumerate}

\item Since $|V_{G'} \setminus S| \leq k$ and $|V_P| \geq k+1$, it holds that $S$ contains at least one purple vertex. Let $v \in S \cap V_P$, since $\deg_S(v) \ge \lceil\gamma(|S|-1)\rceil \ge \lceil\gamma(k'-1)\rceil = d_{\gamma} = \deg_{G'}(v) \ge \deg_S(v)$, it holds that $\deg_{G'}(v) = \deg_S(v)$, i.e. all neighbors of $v$ are in $S$: which implies that $V_R$ and $V_P$ are contained in $S$. Using the same argument on each other purple vertex and since each yellow vertex has a purple neighbor, it holds that $V_Y \subseteq S$. This proves that every non-blue vertex is in $S$.

\item From what precedes, we obtain that $\lceil\gamma(k'-1)\rceil = \lceil\gamma(|S|-1)\rceil$. By Claim~\ref{claim:tech_N}.2 it holds that $\left\lceil \gamma(k'-1) \right\rceil < \left\lceil \gamma k' \right\rceil$. Using $|S|\ge k'$, we get $|S|=k'$.

\item Let $\{u,v\}\in \overline{E_G}$. Assume by contradiction that neither $u$ nor $v$ is in $S$. By what precedes, $\overline{e}_{u,v}\in S$ and since $u$ and $v$ are two distinct neighbors of $\overline{e}_{u,v}$ outside of $S$, we have $\deg_S(\overline{e}_{u,v}) \le \deg_{G'}(\overline{e}_{u,v}) -2 = d_{\gamma} -1 < \lceil \gamma (|S|-1) \rceil$, contradicting that $S$ is a $\gamma$-complete-subgraph. This proves that either $u$ or $v$ belongs to $S$.

\end{enumerate}

\end{proofclaim}

\begin{itemize}
    \item[$\implies$] Let $K \subseteq V_{G}$ be a clique of $G$ of size $k$.

    Let $S = V_{G'} \setminus K$. It holds that $|S| = n+\overline{m}+k+1+N-k = k'$. Also:
    \begin{itemize}
        \item For any blue vertex in $S$ (i.e. $v \in V_{G} \cap S$), let's recall that $v$ is linked to all $V_Y$, so: $\deg_{S}(v) \geq |V_Y| = N \geq d_{\gamma}$.

        \item For any red vertex $\overline{e}_{u,v} \in V_R$, let's recall that $\overline{e}_{u,v}$ is linked in $G$ to $u,v$, and in $S$ to all $k+1$ vertices of $V_P$ and $d_{\gamma}-k-2$ vertices of $V_Y$. Since $u$ and $v$ are not adjacent in $G$ it holds that at least one of them is not in $K$ and thus is in $S$. So $\deg_{S}(\overline{e}_{u,v}) \geq d_{\gamma}-1 + 1 = d_{\gamma}$. 
        
        \item For any purple vertex $v \in V_P$, let's recall that $v$ is linked to all $\overline{m}$ vertices of $V_R$, all $k$ other vertices of $V_P$ and $d_{\gamma}-k-\overline{m}$ yellow vertices. All of its neighbors are in $S$, thus $\deg_{S}(v) = d_{\gamma}$.

        \item For any yellow vertex $v \in V_Y$, let's recall that $v$ is linked to all yellow vertices and at least one purple vertex. Hence, $\deg_{S}(v) \geq N-1+1 \geq d_{\gamma}$.
    \end{itemize}

    Hence, $S$ is a $\gamma$-complete subgraph of $G'$ of size $k'$.

    \item[$\impliedby$] Let $S$ be a $\gamma$-complete subgraph of $G'$ of size $\geq k'$. Let $K = V_{G'} \setminus S$.

    We have $|S|=k'$ by Claim~\ref{claim:gcs_ell}.2, and thus $|K|=|V_{G'}| - |S| = k$. Moreover, $S$ contains every non-blue vertex by Claim~\ref{claim:gcs_ell}.1: it follows that $K$ contains only blue vertices (i.e. vertices of $V_G$). We will prove that $K$ is a clique.

    Let $(u,v)\in K^2$ with $u\neq v$, and assume by contradiction that $\{u,v\}\notin E_G$, i.e. $\{u,v\}\in\overline{E_G}$. By Claim~\ref{claim:gcs_ell}.3, either $u$ or $v$ is in $S$, contradicting that $(u,v)\in K^2$. This proves that $\{u,v\}\in E_G$
    
    Thus $K$ is a clique of $G$ of size $k$.
    
\end{itemize}

\underline{Preservation of the parameter:}

\hfill

Since we reduce the instance $(G,k)$ of \clique{} parameterized by $k$ to the instance $(G',k')$ of \gcs{} parameterized by $\ell$, we need to check that $\ell':= |V_{G'}|-k'$ depends only on $k$. It is indeed true since $\ell'=k$.

\end{proof}

Note that, even if the concept of $s$-plex with $s\ge 2$ (a subset $S\subseteq V_G$ such that $\forall u\in S, \deg_S(u)\ge |S|- s$) is similar to $\gamma$-complete subgraph, this proof cannot be adapted to $s$-plexes. Indeed, we use the fact that with a large enough number of vertices, the vertices can also have a large enough number of non-neighbors in the $\gamma$-complete subgraph. Something similar cannot be said for $s$-plexes where the number of non-neighbors is always constant.

Also, similarly as discussed in the end of Section~\ref{sec:gcs_degeneracy}, unless {\sf P=NP}, the result of {\sf W[1]}-hardness obtained in Theorem~\ref{thm:gCS_W1_l} cannot be improved to a result of para-{\sf NP}-hardness unless {\sf P=NP}. Indeed, testing if every subset $S\subseteq V_G$ with $|S| \geq n-\ell$ is a $\gamma$-complete-subgraph takes time $O(n^2 \times \ell \times \binom{n}{\ell}) = O(n^3\times n^{\ell})$.

\section{Discussion}

In this paper, we have completed the study of the parameterized complexity of \sclub{} when parameterized by the degeneracy $d$, as we established its para-{\sf NP}-hardness for all value of $s\ge 3$. The para-{\sf NP}-hardness of $2$-\club{} was indeed already known \cite{schafer2012parameterized}, and noticing that $1$-\club{}, which corresponds to \clique{}, is {\sf FPT} when parameterized by $d$, all cases are know covered. More precisely, if $s\ge 3$ is odd, we proved that \sclub{} is {\sf NP}-hard even on bipartite graphs of degeneracy $3$, and if $s\ge 4$ is even, \sclub{} is {\sf NP}-hard even on graphs of degeneracy $3$\footnote{\label{foot:degeneracy2}We even get the {\sf NP}-hardness for $d=2$ if $s\ge 5$, and for bipartite graphs if $s$ is odd}. We also proved that the exact same complexity results hold for \sclique{}.

Then, we have established the {\sf W[1]}-hardness of the \gcs{} problem parameterized by 3 different parameters: $k$, $\ell$ and $d$, which ends the study of the parameterized complexity of \gcs{} when parameterized by the most relevant parameters, as proposed by Komusiewicz \cite{komusiewicz2016multivariate}. 

Most of the complexity results are now known about \sclub{}, \sclique{} and \gcs{}, our contribution is summarized in Figure~\ref{fig:table}. Note that, unless {\sf P = NP}, all of our {\sf W[1]}-hardness results can not be improved to para-{\sf NP}-hardness results since {\sf XP} algorithm exist to solve these problems, as discussed at the end of Section~\ref{sec:gcs_degeneracy} and Section~\ref{sec:gcs_l}.

\begin{figure}[!ht]

\scalebox{.9}{

\hspace{-1cm}
    
\begin{tabular}{c c c c c c}
    Problem & $k$ & $\ell$ & $h$ & $d$ \\ \hline
    \clique{} & {\sf W[1]}-h~\cite{downey2013} & {\sf FPT}~\cite{komusiewicz2016multivariate} & {\sf FPT}~\cite{komusiewicz2016multivariate}& {\sf FPT}~\cite{komusiewicz2016multivariate}\\
    $2$-\club{} & {\sf FPT}~\cite{komusiewicz2016multivariate} & {\sf FPT}~\cite{komusiewicz2016multivariate} & {\sf W[1]}-h\cite{hartung2015structural} & {\sf NP}-h for $d=6$~\cite{hartung2015structural} \\
    $s$-\club{} with $s\ge 3$ & {\sf FPT}~\cite{komusiewicz2016multivariate} & {\sf FPT}~\cite{komusiewicz2016multivariate}& ? & \underline{\textcolor{red}{\textbf{{\sf NP}-h for $d = 3$}}}\footnotemark[1] \\
    $2$-\clique{} & {\sf FPT}~\cite{komusiewicz2016multivariate} & {\sf FPT}~\cite{komusiewicz2016multivariate} & ? & ? \\
    \sclique{} with $s\ge 3$ & {\sf FPT}~\cite{komusiewicz2016multivariate} & {\sf FPT}~\cite{komusiewicz2016multivariate} & ? & \underline{\textcolor{red}{\textbf{{\sf NP}-h for $d = 3$}}}\footnotemark[1] \\
    \gcs{} & \textcolor{red}{\textbf{{\sf W[1]}-h}} \underline{\color{red} for any $\gamma$} & \underline{\textcolor{red}{\textbf{{\sf W[1]}-h}}} & {\sf FPT}~\cite{baril2021hardness} & \underline{\textcolor{red}{\textbf{{\sf W[1]}-h}}}  \\
\end{tabular}

}

\caption{Complexities of relaxations of \clique, our contributions underlined.}
    \label{fig:table}
\end{figure}

Nevertheless, the complexity of \sclub{} and \sclique{} parameterized by the $h$-index (even though the particular case of $2$-\club{} is known \cite{schafer2012parameterized}), and the complexity of $2$-{\sc Clique} when parameterized by the degeneracy $d$ remain open. Note that as $1$-clique is {\sf FPT} parameterized by $d$ and as $s$-clique is para-{\sf NP}-hard for the same parameter, this case is intermediate. Moreover, for $s \in \{3,4\}$, the graph obtained by our reduction is only $3$-degenerated and for $s = 2$, the graph obtained \cite{hartung2015structural} is only $6$-degenerated. Thus, the complexity for $2$-degenerated graphs for $s \in \{3,4\}$, and the complexity on the class of $d$-degenerated graphs, for $d \in \{2,...,5\}$ and $s=2$ are also still open .

\newpage

\bibliographystyle{plain}
\bibliography{biblio}

\end{document}